\documentclass[%
 superscriptaddress,
 twocolumn,
 amsmath,amssymb,
 aps,
 prb,
]{revtex4-2}

\usepackage{graphicx}
\usepackage{dcolumn}
\usepackage{bm}
\usepackage{comment}

\usepackage{color}

\begin{document}


\title{Optical non-Hermitian skin effect in two-dimensional uniform media}

\author{Taiki Yoda}
\affiliation{NTT Basic Research Laboratories, NTT Corporation, 3-1 Morinosato-Wakamiya, Atsugi-shi, Kanagawa 243-0198, Japan}

\author{Yuto Moritake}
\affiliation{Department of Physics, Tokyo Institute of Technology, 2-12-1 Ookayama, Meguro-ku, Tokyo 152-8551, Japan}
\affiliation{PRESTO, Japan Science and Technology Agency, 4-1-8 Honcho, Kawaguchi, Saitama 332-0012, Japan}

\author{Kenta Takata}
\affiliation{NTT Basic Research Laboratories, NTT Corporation, 3-1 Morinosato-Wakamiya, Atsugi-shi, Kanagawa 243-0198, Japan}
\affiliation{Nanophotonics Center, NTT Corporation, 3-1, Morinosato-Wakamiya, Atsugi-shi, Kanagawa 243-0198, Japan}

\author{Kazuki Yokomizo}
\affiliation{Department of Physics, The University of Tokyo, Hongo, Bunkyo-Ku, Tokyo 113-0033, Japan}

\author{Shuichi Murakami}
\affiliation{Department of Physics, Tokyo Institute of Technology, 2-12-1 Ookayama, Meguro-ku, Tokyo 152-8551, Japan}

\author{Masaya Notomi}
\affiliation{NTT Basic Research Laboratories, NTT Corporation, 3-1 Morinosato-Wakamiya, Atsugi-shi, Kanagawa 243-0198, Japan}
\affiliation{Department of Physics, Tokyo Institute of Technology, 2-12-1 Ookayama, Meguro-ku, Tokyo 152-8551, Japan}
\affiliation{Nanophotonics Center, NTT Corporation, 3-1, Morinosato-Wakamiya, Atsugi-shi, Kanagawa 243-0198, Japan}

\date{\today}

\begin{abstract}
The non-Hermitian skin effect (NHSE) is a novel localization phenomenon in certain non-Hermitian systems with gain and/or loss.
Most of previous works study the non-Hermitian skin effect in periodic systems.
However, electromagnetic waves often propagate within uniform materials without periodic modulation, and it has not been clear whether the optical NHSE occurs in uniform media such as bulk materials and electromagnetic metamaterials.
Here we establish the theory of the optical NHSE in non-Hermitian anisotropic media.
We show that the NHSE occurs even in uniform media with appropriate anisotropy and material loss.
The localization of non-Hermitian skin modes are completely determined by an effective gauge potential caused by the anisotropy of a dielectric tensor.
On the basis of the theory, we propose subwavelength multilayer metamaterials as a novel platform for the optical NHSE.
We also propose a new concept of stationarily-excited skin modes whose frequencies are forced to be real in non-Hermitian systems.
We find that the NHSE occurs even under the condition that the frequency is forced to be real, which implies that the NHSE we propose is observable under stationary excitation.
Our work presents a general theory of the NHSE in homogeneous systems, and pave the way to realize the optical NHSE in bulk materials and metamaterials.
\end{abstract}

\maketitle

\section{Introduction}
Non-Hermitian systems with gain and/or loss have been extensively investigated because non-Hermiticity often leads to novel phenomena without counterparts in Hermitian systems~\cite{feng2017non,el2018non,ozdemir2019parity,OtaTakataOzawaAmoJiaKanteNotomiArakawaIwamoto+2020+547+567}.
Recent studies have shown that ``bulk" eigenstates are localized at the boundary of systems in certain non-Hermitian periodic systems, and that their localized states form continuous spectra in spite of the localization.
Such peculiar localized phenomenon is called the non-Hermitian skin effect (NHSE)~\cite{PhysRevLett.121.086803,PhysRevLett.121.136802,PhysRevLett.123.066404,PhysRevLett.124.086801,PhysRevB.101.195147}.
Most previous works related to the NHSE have discussed discrete lattice systems described by tight binding models.
However, the propagation of electromagnetic waves in many photonic systems is described by differential equations such as the Maxwell equation.
Therefore, the theory of the NHSE based on discrete tight binding models cannot be applied to general optical materials. 
Recently, several works have investigated the NHSE in periodically-modulated systems beyond tight binding models~\cite{PhysRevB.104.125109,PhysRevB.104.125416,yan2021non,PhysRevResearch.4.023089,FangHuZhouDing+2022+3447+3456,PhysRevB.106.195412}, and discussed an optical NHSE in photonic crystals with periodic structure~\cite{PhysRevB.104.125416,yan2021non,PhysRevResearch.4.023089,FangHuZhouDing+2022+3447+3456,PhysRevB.106.195412}.

This paper has three main purposes.
First, we theoretically show that the NHSE occurs even in uniform media with a position-independent dielectric tensor.
Most of the previous works related to the NHSE focus on periodic systems because the NHSE stems from condensed matter physics.
The NHSE in periodic systems is closely related to the notion of the Brillouin zone and Bloch wavevector~\cite{PhysRevLett.121.086803,PhysRevLett.121.136802,PhysRevLett.123.066404,PhysRevLett.124.086801,PhysRevB.101.195147,PhysRevResearch.4.023089}. 
However, electromagnetic waves often propagate in an optically uniform medium with a uniform dielectric permittivity such as bulk materials and metamaterials.
Therefore, in the field of optics (and other classical wave systems), it is significant whether the NHSE occurs even in uniform systems without periodic modulation.
Inspired by Ref.~\cite{PhysRevResearch.4.023089}, we theoretically study two-dimensional uniform anisotropic media with a non-Hermitian dielectric tensor, and successfully derive the analytical solution of the NHSE in non-Hermitian anisotropic media.
Although absorption of electromagnetic waves in lossy anisotropic materials was investigated~\cite{PhysRevB.86.195411,nefedov2013total,Nefedov_2013,Debnath:19}, the NHSE in lossy (or gainy) anisotropic materials has not been pointed out to our knowledge.

Second, we propose multilayer metamaterials as a novel platform for the optical NHSE.
Although the optical NHSE in elecromagnetic metamaterials has not been proposed, our theory enable us to realize the optical NHSE in metamaterials.
We can implement lossy anisotropic systems by using sub-wavelength metamaterials because they can be regarded as anisotropic media when they satisfy the effective medium condition.
We particularly consider multilayer metamaterials consisting of alternating metal and dielectric layers because they exhibit strong in-plane anisotropy~\cite{poddubny2013hyperbolic,narimanov2015naturally}.
Multilayer metamaterials will be good candidate for the optical NHSE.

Third, we propose a new concept of stationarily-excited skin modes in two-dimensional systems.
If we consider observation of NHSE, we have to take excitation processes into account. However, it is not guaranteed that eigenstates are excited and observed in non-Hermitian systems, especially for NHSE, and it is not generally trivial how to account excitation processes. Our analytical framework of the NHSE enables us to solve this problem rather directly.
In the previous research of NHSE in two-dimensional systems~\cite{PhysRevLett.121.136802,PhysRevB.104.125416,FangHuZhouDing+2022+3447+3456,PhysRevResearch.2.023265,PhysRevResearch.2.022062,PhysRevLett.125.118001,PhysRevB.102.241202,PhysRevB.103.205205,zhang2021observation,zhang2022universal,yokomizo2022non}, the localization of eigenmodes with generally complex eigenvalues was investigated.
However, we can also choose a solution with a real frequency and a complex wavevector in non-Hermitian systems~\cite{doi:10.1021/acsphotonics.9b01202}.
The concept of a stationarily-excited skin mode is motivated by realistic optical experiments such as transmission and reflection measurement: we often examine the response of an optical system under stationary excitation, where the frequency of excited modes is forced to be real.
It has been not clear that the NHSE occurs under the the condition that frequency is forced to be real.
By using the analytical solution of the NHSE in uniform media, we show that a real-$\omega$ mode also exhibits the NHSE as well as a real-$k$ mode.
The concept of a stationarily-excited skin mode allows us to calculate the propagation length in a propagation direction under the real-$\omega$ condition, in addition to the localization length perpendicular to the propagation direction.
More importantly, the localization length of a real-$\omega$ skin mode generally differs from that of a real-$k$ skin mode.
The distinction between a real-$k$ skin mode and a real-$\omega$ skin mode is crucial because one may exhibit NHSE but the other may not exhibit NHSE.
We must appropriately choose the value of frequency and wavevector depending on the situation and how to excitation.

\section{Eigenmode analysis of non-Hermitian anisotropic media}
\label{sec:2}
\subsection{Theory}
\label{sec:real-k_theory}
The propagation of electromagnetic waves in an uniform medium is governed by a dielectric tensor.
The energy of electromagnetic waves is generally not conserved in a system with a non-Hermitian dielectric tensor~\cite{landau2013electrodynamics}.
In this paper, we focus on uniform media with a non-Hermitian dielectric tensor given by
\begin{align}
\begin{pmatrix}
\varepsilon_{xx} && \varepsilon_{xy} && 0
\\
\varepsilon_{yx} && \varepsilon_{yy} && 0
\\
0 && 0 && \varepsilon_{zz}
\end{pmatrix}.
\label{eq:epsilon}
\end{align}
We denote the upper-left block of Eq.~(\ref{eq:epsilon}) as $\varepsilon$.
The magnetic permeability is assumed to be unity.
The dielectric tensor could be reciprocal or non-reciprocal. We will discuss the effect of reciprocity later.
When electromagnetic waves propagate within the $xy$ plane in the media, transverse electric (TE) modes and transverse magnetic modes are decoupled each other.
The uniform solution of TE modes in the $y$ direction, which is expressed as $H_{z}(x, y, t) = e^{i(\omega t - k_{y}y)}H_{z}(x)$, satisfies the following equation:~\cite{PhysRevResearch.4.023089}
\begin{gather}
\hat{\Theta}\left( k_{y} \right)H_{z} \left( x \right) = \left( \frac{\omega}{c} \right)^{2}H_{z} \left( x \right),
\label{eq:eigenequation}
\\
\hat{\Theta} \left( k_{y} \right) =
-\eta_{yy}\frac{d^{2}}{dx^{2}} - ik_{y} \left( \eta_{xy} + \eta_{yx} \right) \frac{d}{dx} + \eta_{xx}k_{y}^{2},
\label{eq:definition_theta}
\\
\begin{pmatrix}
\eta_{xx} && \eta_{xy}
\\
\eta_{yx} && \eta_{yy}
\end{pmatrix}
= \frac{1}{\varepsilon_{xx}\varepsilon_{yy} - \varepsilon_{xy}\varepsilon_{yx}}
\begin{pmatrix}
\varepsilon_{yy} && -\varepsilon_{xy}
\\
-\varepsilon_{yx} && \varepsilon_{xx}
\end{pmatrix},
\end{gather}
where $\omega$ is the angular frequency, $k_{y}$ is the $y$ component of the wavevector, and $c$ is the speed of light in vacuum.
Equations (\ref{eq:eigenequation}) and (\ref{eq:definition_theta}) take the similar form to the one-dimensional Schr\"{o}dinger equation with a gauge potential.
The non-Hermiticity of $\varepsilon$ is a necessary condition for the non-Hermiticity of $\hat{\Theta}(k_{y})$.
The second term in Eq.~(\ref{eq:definition_theta}) corresponds to an effective gauge potential for photon induced by the anisotropy of a dielectric tensor~\cite{PhysRevLett.114.103902,chen2019non,PhysRevResearch.4.023089,PhysRevLett.124.086801,PhysRevB.101.195147,brandenbourger2019non}.
Therefore, electromagnetic waves in anisotropic media can potentially emulate the dynamics of free electrons in a uniform gauge potential.
The first derivative in Eq.~(\ref{eq:definition_theta}) gives spatial asymmetry of a system along the $x$ direction, and it can be effectively regarded as asymmetric hopping in tight binding models~\cite{PhysRevLett.121.086803,PhysRevLett.123.066404}.
Although Eqs.~(\ref{eq:eigenequation}) and (\ref{eq:definition_theta}) are similar to Eq.~(5) of Ref.~\cite{PhysRevResearch.4.023089}, we deal with position-independent dielectric tensors in this paper.
We note that the first derivative term in Eq.~(\ref{eq:definition_theta}) vanishes when $k_{y} = 0$.
Therefore, the $y$-dependence of $H_{z}$ is crucial for achieving an effective gauge potential, and NHSE caused by the anisotropy of a dielectric tensor is essentially a two-dimensional phenomenon.

The anti-Hermitian part of an effective gauge potential causes peculiar localization of eigenmodes like electrons in an imaginary gauge potential~\cite{PhysRevLett.77.570}.
To see this, we consider uniform systems which are finite in the $x$ direction with a size $L$ and infinite in the $y$ direction.
As was clarified previously, the eigenfrequency and mode profile are sensitive to boundary conditions in non-Hermitian systems. Eigenmodes are always extended for the periodic boundary condition (PBC) having an infinite medium size, but localized skin modes generally appear for finite-sized boundary conditions.
We first impose the periodic boundary condition (PBC) $H_{z}(0) = H_{z}(L)$.
The PBC quantizes the wavevector $k_{x}$ to $2\pi n/L$, and we obtain
\begin{align}
k_{x} = \frac{2\pi n}{L}, \quad n = 0, \pm 1, \pm 2,\cdots,
\label{eq:PBC-kx}
\end{align}
\begin{align}
\left( \frac{\omega_{\text{PBC}}}{c} \right)^{2}
&= \eta_{yy}k_{x}^{2} - \left( \eta_{xy} + \eta_{yx} \right) k_{x}k_{y} + \eta_{xx}k_{y}^{2},
\nonumber \\
&= \eta_{yy} \left( k_{x} + qk_{y} \right)^{2} + \left( \eta_{xx} - q^{2}\eta_{yy} \right)k_{y}^{2},
\label{eq:PBC-omega}
\\
q &= -\frac{\eta_{xy} + \eta_{yx}}{2\eta_{yy}} = \frac{\varepsilon_{xy} + \varepsilon_{yx}}{2\varepsilon_{xx}},
\label{eq:definition_q}
\end{align}
where we define the dimensionless parameter $q$.
Equation (\ref{eq:PBC-omega}) is the dispersion relation of a planewave solution with a real $k_{x}$.
It follows from Eq.~(\ref{eq:PBC-omega}) that $qk_{y}$ corresponds to the effective gauge potential because it shifts the wavevector $k_{x}$.
The parameter $q$ is an important quantity in our work, which essentially determine the strength of the effective gauge potential for each medium.
By definition, the effective gauge potential can be drastically enhanced when $|\varepsilon_{xx}| \approx 0$.

Under the PBC, the wavevector $k_{x}$ becomes real and thus the electromagnetic wave is extended over the system.
However, the delocalization of the eigenmode does not necessarily hold under other boundary conditions.
The general solution of Eq.~(\ref{eq:eigenequation}) is expressed by the superposition of two planewaves:
\begin{gather}
\begin{cases}
H_{z}(x)
= Ae^{-ik_{-}x} + Be^{-ik_{+}x},
\\
E_{x}(x) = -\left( Z_{y-}Ae^{-ik_{-}x} + Z_{y+}Be^{-ik_{+}x} \right),
\\
E_{y}(x) = Z_{x-}Ae^{-ik_{-}}x + Z_{x+}Be^{-ik_{+}x},
\end{cases}
\label{eq:general-solution}
\end{gather}
where $k_{\pm}$, $Z_{x\pm}$, and $Z_{y\pm}$ are defined by
\begin{align}
Z_{y\pm} &= \frac{1}{\omega\varepsilon_{0}} \left( k_{y}\eta_{xx} - k_{\pm}\eta_{xy} \right),
\\
Z_{x\pm} &= \frac{1}{\omega\varepsilon_{0}} \left( -k_{y}\eta_{yx} + k_{\pm}\eta_{yy} \right),
\\
k_{\pm} &= -qk_{y} \pm \alpha,
\label{eq:wavevector}
\\
\alpha &= \sqrt{ \frac{1}{\eta_{yy}} \left\{ \left( \frac{\omega}{c} \right)^{2} - \left( \eta_{xx} - q^{2}\eta_{yy} \right)k_{y}^{2} \right\} },
\end{align}
where $\varepsilon_{0}$ is the permittivity of vacuum.
The two wavevectors satisfy $k_{-} \neq -k_{+}$ when $q \neq 0$~\cite{PhysRevB.86.195411,nefedov2013total,Nefedov_2013,Debnath:19}.
The value of $k_{\pm}$ is determined by boundary conditions.
As a specific example, let us consider a system sandwiched by two perfect electric conductors (PECs) placed at $x = 0$ and $L$.
The two PECs require $E_{y}(0) = E_{y}(L) = 0$, which quantizes the difference of the wavevectors as $k_{+} - k_{-} = 2\alpha = 2\pi n/L$.
By combining $k_{+} + k_{-} = -2qk_{y}$, we obtain
\begin{gather}
k_{\pm} = \pm\frac{\pi n}{L} - qk_{y}, \quad n = 1,2,\cdots.
\label{eq:PEC-wavevector}
\end{gather}
Inserting Eq.~(\ref{eq:PEC-wavevector}) into Eqs.~(\ref{eq:general-solution}) and (\ref{eq:wavevector}) yields eigenfrequency $\omega_{\text{PEC}}$ and eigenmode,
\begin{align}
\left( \frac{\omega_{\text{PEC}}}{c} \right)^{2}
= \eta_{yy} \left( \frac{\pi n}{L} \right)^{2} + \left( \eta_{xx} - q^{2}\eta_{yy} \right) k_{y}^{2},
\label{eq:PEC-eigenfrequency}
\end{align}
\begin{align}
E_{y}(x) &= e^{iqk_{y}x}\sin \left( \frac{\pi n}{L}x \right),
\label{eq:PEC-eigemode-Ey}
\\
H_{z}(x) &= \frac{1}{2}ie^{iqk_{y}x}
\Biggl[
\left( \frac{1}{Z_{x+}} - \frac{1}{Z_{x-}} \right) \cos\left( \frac{\pi n}{L}x \right)
\nonumber \\
&- i \left( \frac{1}{Z_{x+}} + \frac{1}{Z_{x-}} \right) \sin\left( \frac{\pi n}{L}x \right)
\Biggr],
\label{eq:PEC-eigemode-Hz}
\\
E_{x}(x) &= -\frac{1}{2}ie^{iqk_{y}x}
\Biggl[
\left( \frac{Z_{y+}}{Z_{x+}} - \frac{Z_{y-}}{Z_{x-}} \right) \cos\left( \frac{\pi n}{L}x \right)
\nonumber \\
&- i \left( \frac{Z_{y+}}{Z_{x+}} + \frac{Z_{y-}}{Z_{x-}} \right) \sin\left( \frac{\pi n}{L}x \right)
\Biggr],
\label{eq:PEC-eigemode-Ex}
\end{align}
Equations (\ref{eq:PEC-wavevector})-(\ref{eq:PEC-eigemode-Ex}) are one of the main results of our work.
First, Eq.~(\ref{eq:PEC-wavevector}) shows that the wavevectors $k_{\pm}$ are shifted from $\pm \pi n/L$ by the effective gauge potential, and that necessarily become complex when $\text{Im}(qk_{y}) \neq 0$.
The imaginary part of $k_{\pm}$ is given by $\text{Im}(k_{+}) = \text{Im}(k_{-}) = -\text{Im}(qk_{y})$.
Second, Eqs.~(\ref{eq:PEC-eigemode-Ey})-(\ref{eq:PEC-eigemode-Ex}) show that all the eigenmodes are localized at a boundary of the system when $\text{Im}(qk_{y}) \neq 0$ because the amplitude of $E_{y}$ is written as $|E_{y}(x)| = \text{exp}[-\text{Im}(qk_{y})x] |\sin(\pi nx/L)|$.
The envelope of the eigenmode decays exponentially and its localization strength is given by $|\text{Im}(k_{\pm})| = |\text{Im}(qk_{y})|$.
The sign of $\text{Im}(qk_{y})$ determines which side the eigenmode is localized at.
Third, the trajectory of complex $\omega_{\text{PEC}}$ (Eq.~(\ref{eq:PEC-eigenfrequency})) in the complex eigenfrequency plane disagrees with $\omega_{\text{PBC}}$ (Eq.~(\ref{eq:PBC-omega})) even in the limit of $L \rightarrow \infty$ when $\text{Im}(qk_{y}) \neq 0$.
As proven in Appendix \ref{sec:PBC-PEC}, the trajectory of $\omega_{\text{PEC}}^{2}$ always is a semi-infinite line on the complex-$\omega^{2}$ plane, while the trajectory of $\omega_{\text{PBC}}^{2}$ is a parabola when $\text{Im}(qk_{y}) \neq 0$.
The two trajectories coincide with each other when $\text{Im}(qk_{y}) = 0$.
We will show numerical examples of these results in Sec.~\ref{sec:reak-k_numerical}.
These results hold for a perfect magnetic conductor condition (see Appendix \ref{sec:PMC}).
The localization will appear for other open boundary conditions such as air-cladding although the analytical expression of its localization length cannot be derived~\cite{PhysRevB.104.125416,PhysRevB.106.195412}.

The localization phenomenon discussed above has the same features as the NHSE in periodic systems.
For the NHSE in periodic systems, the localization of a skin mode is described by the non-Bloch band theory~\cite{PhysRevLett.121.086803,PhysRevLett.121.136802,PhysRevLett.123.066404,PhysRevB.101.195147,yan2021non,PhysRevResearch.4.023089,yokomizo2022non}.
The non-Bloch band theory extend the Bloch wavevector to the complex number, which corresponds to Eq.~(\ref{eq:PEC-wavevector}).
The imaginary part of a Bloch wavevector determines the localization strength and localized position of a skin mode.
Although such explanations may seem valid only for periodic systems, essentially the same explanations are valid in our uniform systems when we replace a Bloch wavevector with a wavevectors as shown in Eq.~(\ref{eq:PEC-wavevector}) and Eqs.~(\ref{eq:PEC-eigemode-Ey})-(\ref{eq:PEC-eigemode-Ex}).
In addition, Eq.~(\ref{eq:PEC-wavevector}) and Eqs.~(\ref{eq:PEC-eigemode-Ey})-(\ref{eq:PEC-eigemode-Ex}) are natural extension of the result in Ref.~\cite{PhysRevResearch.4.023089}:
in one-dimensional periodic crystals, the decay of a skin mode is determined by the imaginary part of the unit-cell integral of a gauge potential.
For the NHSE in periodic systems, the eigenvalue under the PBC disagrees with that under an open boundary condition, which corresponds to Eq.~(\ref{eq:PBC-omega}) and (\ref{eq:PEC-eigenfrequency}).
Consequently, the localization phenomenon described by Eqs.~(\ref{eq:PEC-eigemode-Ey})-(\ref{eq:PEC-eigemode-Ex}) are considered as the NHSE in uniform media.

Despite of similarity presented above, we point out a few issues which contrasts NHSE in periodic and uniform media.
First, the fact that eigenmodes of the NHSE are completely described by an analytical form is very important. Various complicated aspects of the NHSE can be analytically examined and classified. We will pursue this aspect in the following parts of this paper. As a first example, we examine the reality condition of $\omega$ using the analytical framework.
The eigenfrequency $\omega_{\text{PEC}}$ given by (\ref{eq:PEC-eigenfrequency}) is generally complex.
The imaginary part of $\omega$ determines whether the eigenmode is attenuated $(\text{Im}(\omega)>0)$ or amplified $(\text{Im}(\omega)<0)$ in time.
When $\text{Im}(\eta_{yy}) > 0$ and $\text{Im}(\eta_{xx} - q^{2}\eta_{yy}) > 0$, $\omega_{\text{PEC}}$ satisfies $\text{Im}(\omega_{\text{PEC}}) > 0$ for all $n$ and $k_{y}$ (when $k_{y}$ is assumed to be real).
Similarly, $\omega_{\text{PEC}}$ satisfies $\text{Im}(\omega_{\text{PEC}}) < 0$ for all $n$ and $k_{y}$ when $\text{Im}(\eta_{yy}) < 0$ and $\text{Im}(\eta_{xx} - q^{2}\eta_{yy}) < 0$.
When $\text{Im}(\eta_{yy}) = 0$ and $\text{Im}(\eta_{xx} - q^{2}\eta_{yy}) = 0$, the reality of $\omega^{2}_{\text{PEC}}$ is guaranteed for all $n$ and $k_{y}$.
The mirror-time symmetry of $\varepsilon$ sufficiently guarantees the reality of $\omega^{2}_{\text{PEC}}$ (see Fig.~\ref{fig:eigen}(g) and \ref{fig:eigen}(h)).
The mirror-time symmetry holds when the diagonal components are real and off-diagonal components are pure imaginary.
The detailed discussion of the mirror-time symmetry is given in Appendix \ref{sec:mirror-time}.

Next, we examine the topological property.
In contrast to periodic systems, the notion of the Brillouin zone cannot be applied to uniform systems.
The real part of the Bloch wavevector is bounded within the first Brillouin zone and it forms a closed loop in the momentum space, while the real part of $k_{\pm}$ is not bounded.
For the NHSE in periodic systems, the eigenvalue under the PBC forms a finite or infinite number of closed loops in the complex eigenvalue plane, and the winding number can be defined for each closed loop~\cite{PhysRevLett.124.086801}.
For the NHSE in uniform media, the eigenvalue under the PBC forms an open arc in the complex eigenvalue space because the real part of $k_{\pm}$ itself does not forms a closed loop in the momentum space.
Nevertheless, we can prove that the eigenfrequency of a skin mode appears inside an open arc drawn by the trajectory of $\omega_{\text{PBC}}$, and an open arc can be characterized by the winding number~\cite{PhysRevB.104.125109}
\begin{align}
W \left( k_{y}, \omega_{0} \right) = \int_{-\infty}^{\infty} \frac{dk_{x}}{2\pi} \text{arg} \left[ \omega_{\text{PBC}}^{2} \left( k_{x}, k_{y} \right) - \omega_{0}^{2} \right],
\end{align}
where $\omega_{0}^{2}$ is a reference point on the complex-$\omega^{2}$ plane.
The winding number $W$ is finite when $\omega_{0}^{2}$ is inside an open arc, and its value is $-\text{sgn}[\text{Im}(qk_{y})]$.
The proof is given in Appendix \ref{sec:winding}.
The direction of a parabola and the sign of the winding number are closely related to the localized position of a skin mode because they are given by the sign of $\text{Im}(qk_{y})$.

The skin mode in uniform media is qualitatively different from surface localized waves in uniform media such as surface plasmon polaritons (SPPs).
For a given $k_{y}$, a SPP has discrete spectrum while a skin mode forms continuous spectrum in the limit of $L \rightarrow \infty$.
For SPPs in lossless isotropic media, the localization length is determined by $\text{Im}(\alpha) = \text{Im}\sqrt{\varepsilon(\omega/c)^{2} - k_{y}^{2}}$.
On the other hand, the localization length of a skin mode is determined by the difference of two wavevectors $\text{Im}(qk_{y})$.
Finally, the localization side of a skin mode in uniform media depends on the propagation direction.
If a skin mode propagating toward $+k_{y}$ is localized at the right (left) side of a system, a skin mode propagating toward $-k_{y}$ is localized at the left (right) side of a system~\cite{PhysRevB.104.125416,PhysRevResearch.4.023089,FangHuZhouDing+2022+3447+3456,PhysRevResearch.2.023265}.

\subsection{Classification of dielectric tensor}
\label{sec:classification}
The existence of the NHSE in uniform media is basically governed by the gauge potential parameter $q$. If $q$ has an imaginary component, this potentially leads to the NHSE. Equation (\ref{eq:definition_q}) tells us that this condition is fully determined by the dielectric tensor, and might be satisfied for a wide variety of anisotropic media with gain or loss. Hence, here we investigate the dielectric tensor and clarify in which class of non-Hermitian anisotropic uniform media the NHSE cannot occur.

First of all, it is clear from the definition of $q$ that the NHSE vanishes when $\varepsilon$ is diagonal because of $q=0$.
This means the NHSE requires some type of anisotropy, in other words, some class of symmetry should be broken. 
It is also obvious that the NHSE cannot occur when $\varepsilon$ is anti-symmetric.
Anti-symmetric dielectric tensors generally appear for materials exhibiting simple circular dichroism or magneto-optical effects. Thus, NHSE cannot occur for materials showing circular dichroism or magneto-optical effect without further anisotropy.

We next discuss two internal symmetries: Lorentz reciprocity and time-reversal symmetry.
Non-Hermitian systems can be classified into three classes according to the Lorentz reciprocity and time-reversal symmetry~\cite{PhysRevA.99.033839,PhysRevLett.124.257403,PhysRevA.105.023509}: reciprocal systems without time-reversal symmetry, non-reciprocal systems with time-reversal symmetry, and non-reciprocal systems without time-reversal symmetry.
Non-Hermitian systems with the Lorentz reciprocity are described by complex symmetric tensors with $\varepsilon_{yx} = \varepsilon_{xy}$.
The presence of the reciprocity simply modifies the definition of $q$ as $q = \varepsilon_{xy}/\varepsilon_{xx}$, and reduces $\eta_{xx} - q^{2}\eta_{yy}$ to $1/\varepsilon_{xx}$.
Therefore, the NHSE can occur in uniform reciprocal media if $\varepsilon_{xy}/\varepsilon_{xx}$ has non-zero imaginary component. In the case of non-reciprocal media, the NHSE can still occur when $(\varepsilon_{xy} + \varepsilon_{yx})/2\varepsilon_{xx}$ has non-zero imaginary component. Note that breaking the Lorentz reciprocity~\cite{jalas2013and,asadchy2020tutorial} is not required for realization of the present NHSE in uniform media. This contrasts with most examples of the NHSE in discrete systems which employ non-reciprocal hopping, but reciprocal skin effects have been reported in Ref.~\cite{PhysRevB.104.125416,PhysRevResearch.4.023089,FangHuZhouDing+2022+3447+3456,PhysRevResearch.2.023265}.
Non-Hermitian systems with the time-reversal symmetry are described by real non-symmetric tensors~\cite{PhysRevLett.124.257403}.
When $\varepsilon$ is a real matrix, $q$ is also real. Thus, the eigenmode in non-Hermitian and time-reversal uniform media do not exhibit the NHSE regardless of its anisotropy.

In Sec.~\ref{sec:real-k_theory}, $q$ is represented by the $xy$ components of the dielectric tensor.
Here we will associate $q$ with the eigenvalue and eigenpolarizations of the dielectric tensor.
We limit ourselves to dielectric tensors described by non-Hermitian normal matrices because a normal matrix is diagonalizable and the eigenvector of a normal matrix forms
an orthogonal basis.
We note that the definition (\ref{eq:definition_q}) itself is valid even when the dielectric tensor is non-normal.
The dielectric tensor $\varepsilon$ is normal when it satisfies
\begin{align}
\left| \varepsilon_{xy} \right| &= \left| \varepsilon_{yx} \right|,
\\
\varepsilon_{xy}^{*} \left( \varepsilon_{xx} - \varepsilon_{yy} \right) &= \varepsilon_{yx}\left( \varepsilon_{xx}^{*} - \varepsilon_{yy}^{*} \right).
\end{align}
It is convenient to express a non-Hermitian normal matrix by
\begin{align}
\varepsilon = aI + b \left[ \sin\psi \cos\delta \sigma_{x} + \sin\psi \sin\delta \sigma_{y} + \cos\psi \sigma_{z} \right],
\label{eq:normal}
\end{align}
where $0 \leq \psi \leq \pi$, $-\pi < \delta \leq \pi$, $\sigma_{i}$ are the Pauli matrices, $a = (1/2)(\varepsilon_{xx} + \varepsilon_{yy})$ and $b$ is a complex number.
Equation (\ref{eq:normal}) becomes diagonal matrices when $\psi = 0$ or $\pi$, becomes symmetric matrices when $\delta = 0$ or $\pi$, and becomes anti-symmetric matrices when $\delta = \pm \pi/2$.
The eigenvalues $\varepsilon_{1,2}$ and corresponding eigenvectors $u_{1,2}$ of Eq.~(\ref{eq:normal}) are given by
\begin{align}
\varepsilon_{1,2} &= a \pm b,
\\
u_{1} &= \left( \cos\frac{\psi}{2}, e^{i\delta}\sin\frac{\psi}{2} \right)^{T},
\\
u_{2} &= \left( -e^{-i\delta}\sin\frac{\psi}{2}, \cos\frac{\psi}{2} \right)^{T}.
\end{align}
The two eigenvectors generally describe two orthogonal elliptical polarizations.
The angle of the long axis of the two elliptical polarization, denoted by $\phi$, is given by $\tan 2\phi = \tan \psi \cos\delta$~\cite{yariv2007photonics}.

We particularly discuss the eigenvector of symmetric tensors and anti-symmetric tensors.
Non-Hermitian systems with the Lorentz reciprocity are described by complex symmetric tensors.
Non-Hermitian symmetric tensors can be derive by putting $\delta = 0$ or $\delta = \pi$ in Eq.~(\ref{eq:normal}).
The two eigenvectors are given by
\begin{align}
u_{1} &= \left( \cos\frac{\psi}{2}, \pm \sin\frac{\psi}{2} \right)^{T},
\label{eq:eigenvector_symmetric1}
\\
u_{2} &= \left( \mp \sin\frac{\psi}{2}, \cos\frac{\psi}{2} \right)^{T}.
\label{eq:eigenvector_symmetric2}
\end{align}
The normal symmetric tensor describes systems where two eigenpolarizations are two orthgonal linear polarizations whose angle is $\psi/2$.
Particularly, Eqs.~(\ref{eq:eigenvector_symmetric1}) and (\ref{eq:eigenvector_symmetric2}) represent linearly $x$- and $y$-polarized when $\psi = 0$ or $\psi = \pi$.
Next, we discuss normal anti-symmetric tensors because the NHSE in uniform media does not occurs in systems with anti-symmetric tensors.
Non-Hermitian anti-symmetric tensors can be derived by putting $\delta = \pm \pi/2$ in Eq.~(\ref{eq:normal}).
The two eigenvectors are given by
\begin{align}
u_{1} &= \left( \cos\frac{\psi}{2}, \pm i\sin\frac{\psi}{2} \right)^{T},
\label{eq:eigenvector_anti-symmetric1}
\\
u_{2} &= \left( \mp i \sin\frac{\psi}{2}, \cos\frac{\psi}{2} \right)^{T}.
\label{eq:eigenvector_anti-symmetric2}
\end{align}
The normal anti-symmetric tensor describes systems where two eigenpolarizations are orthogonal elliptical polarizations.
The long axis of one elliptocal polarization is oriented along the $x$ direction, and the long axis of the other is oriented along the $y$ direction.
Eigenpolarizations (\ref{eq:eigenvector_anti-symmetric1}) and (\ref{eq:eigenvector_anti-symmetric2}) particularly become two opposite circular polarizations when $\psi = \pi/2$.

A normal matrix is diagonalizable by unitary matrices.
By using the unitary transformation, $\varepsilon$ can be expressed by using $\varepsilon_{1,2}, \psi$, and $\delta$,
\begin{align}
\varepsilon &= U
\begin{pmatrix}
    \varepsilon_{1} && 0
    \\
    0 && \varepsilon_{2}
\end{pmatrix}
U^{-1}
\nonumber \\
&=
\begin{pmatrix}
    \varepsilon_{1}\cos^{2}\frac{\psi}{2} + \varepsilon_{2}\sin^{2}\frac{\psi}{2} && (\varepsilon_{1} - \varepsilon_{2}) e^{-i\delta}\cos\frac{\psi}{2}\sin\frac{\psi}{2}
    \\
    (\varepsilon_{1} - \varepsilon_{2}) e^{i\delta}\cos\frac{\psi}{2}\sin\frac{\psi}{2} && \varepsilon_{1}\sin^{2}\frac{\psi}{2} + \varepsilon_{2}\cos^{2}\frac{\psi}{2}
\end{pmatrix},
\\
U &=
\begin{pmatrix}
    \cos\frac{\psi}{2} && -e^{-i\delta}\sin\frac{\psi}{2}
    \\
    e^{i\delta}\sin\frac{\psi}{2} && \cos\frac{\psi}{2}
\end{pmatrix}.
\end{align}
The corresponding $q$ is calculated as
\begin{align}
q = \frac{(\varepsilon_{1} - \varepsilon_{2})\cos\frac{\psi}{2}\sin\frac{\psi}{2}\cos\delta}{\varepsilon_{1}\cos^{2}\frac{\psi}{2}+\varepsilon_{2}\sin^{2}\frac{\psi}{2}}.
\label{eq:definition_q_2}
\end{align}
Using this result, we summarize that NHSE cannot occur when one of the following conditions are satisfied: (1) $\psi = 0$ or $\pi$. (2) $\delta = \pm \pi/2$. (3) $\varepsilon_{1} = \varepsilon_{2}$.
Conditions (1) and (2) means that $q=0$ when the long axis of two polarizations are oriented along the $x$ and $y$ direction.
Condition (3) means that $q=0$ when $\varepsilon$ is isotropic.
Therefore, the optical NHSE generally occurs in anisotropic media with gain or loss, including uniaxially or biaxially anisotropic crystals, when the anisotropy does not fall into these three cases. As pointed out before, anisotropy described by an anti-symmetric tensor, such as circular dichroism or Faraday effect does not lead to NHSE because it satisfies (2), though materials showing circular dichroism or magneto-optic effect can exhibit NHSE when they possess extra anisotropy destroying some of three conditions.

Finally, we separately calculate the real and imaginary parts of Eq.~(\ref{eq:definition_q_2}).
They are given by
\begin{align}
&\text{Re}(q)
= \frac{\cos\frac{\psi}{2}\sin\frac{\psi}{2}\cos\delta}{|\varepsilon_{1}\cos^{2}\frac{\psi}{2} + \varepsilon_{2}\sin^{2}\frac{\psi}{2}|^{2}} \times
\nonumber \\
&\Biggl[
\left\{ |\varepsilon_{1}|^{2} - \text{Re}(\varepsilon_{1}^{*}\varepsilon_{2}) \right\} \cos^{2}\frac{\psi}{2}
- \left\{ |\varepsilon_{2}|^{2} - \text{Re}(\varepsilon_{1}^{*}\varepsilon_{2}) \right\} \sin^{2}\frac{\psi}{2} \Biggr],
\label{eq:Req}
\end{align}
\begin{align}
\text{Im}(q)
= \frac{\cos\frac{\psi}{2}\sin\frac{\psi}{2}\cos\delta}{|\varepsilon_{1}\cos^{2}\frac{\psi}{2} + \varepsilon_{2}\sin^{2}\frac{\psi}{2}|^{2}} \text{Im}(\varepsilon_{1}\varepsilon_{2}^{*}).
\label{eq:Imq_eigenpolarization}
\end{align}
The real and imaginary parts are generally finite.
However, they sometimes accidentally vanishes.
Equation (\ref{eq:Req}) vanishes when the terms in the square brackets vanishes (see Fig.~\ref{fig:eigen}(g) and \ref{fig:eigen}(h)).
Equation (\ref{eq:Imq_eigenpolarization}) vanishes when $\text{arg}(\varepsilon_{1}) = \text{arg}(\varepsilon_{2})$ (mod $\pi$).
So far, we did not assume $k_{y}$ to be real.
In the following section in \ref{sec:2} and \ref{sec:3}, however, we assume real $k_{y}$, which is the normal assumption for eigenmodes of NHSE.
For this case, NHSE is inhibited when $\text{Im}(q)=0$.
Equation (\ref{eq:Imq_eigenpolarization}) means that there appears an additional inhibition condition of NHSE, that is, $\text{arg}(\epsilon_{1}) = \text{arg}(\epsilon_{2})$ (mod $\pi$).
In other words, when the real and imaginary parts of the dielectric ellipsold has the similar anisotropy, such anisotropy does not lead to NHSE.
An example of this case is shown in Fig.~\ref{fig:eigen}(e) and \ref{fig:eigen}(f).
Later in the section \ref{sec:4}, we explicitly deal with modes with complex $k_{y}$.



\subsection{Numerical calculation of analytical result}
\label{sec:reak-k_numerical}
\begin{figure*}
\includegraphics[width=17.2cm]{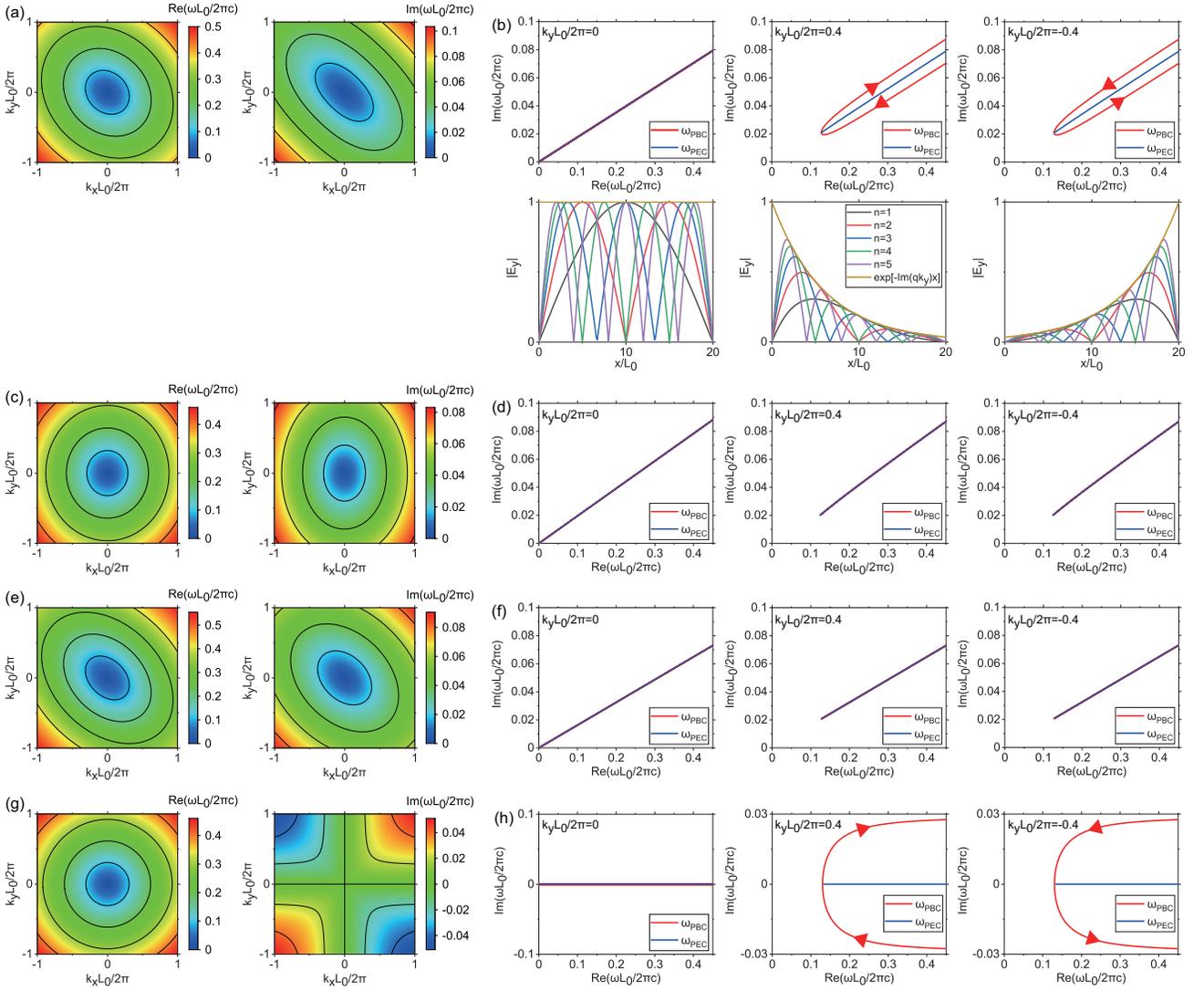}
\caption{
(a)(c)(e)(g) Calculated (left) $\text{Re}(\omega_{\text{PBC}})$ and (right) $\text{Im}(\omega_{\text{PBC}})$ in the limit of $L \rightarrow \infty$ as functions of $k_{x}$ and $k_{y}$, where $L_{0}$ is an arbitrary unit length.
$\omega_{\text{PBC}}$ is calculated from Eq.~(\ref{eq:PBC-omega}).
(b)(d)(f)(h) Calculated complex $\omega_{\text{PBC}}$ and $\omega_{\text{PEC}}$ at (left) $k_{y}L_{0}/2\pi = 0$, (middle) $0.4$, and (right) $-0.4$ on the complex-$\omega$ plane.
$\omega_{\text{PEC}}$ is calculated from Eq.~(\ref{eq:PEC-eigenfrequency}).
Red arrows represent the direction of the trajectory when $k_{x}$ increases.
Lower panels of (b) plots $|E_{y}|$ calculated from Eq.~(\ref{eq:PEC-eigemode-Ey}) when $k_{y}L_{0}/2\pi = 0$ and $k_{y}L_{0}/2\pi = \pm0.4$.
In the lower panels of (b), we set $L=20L_{0}$.
The used value of the dielectric tensor is as follows: (a)(b) $\varepsilon_{xx} = \varepsilon_{yy} = 9 - 3i$ and $\varepsilon_{xy} = \varepsilon_{yx} = 2$.
(c)(d) $\varepsilon_{xx} = 10 - 3i$, $\varepsilon_{yy} = 8 - 3i$, $\varepsilon_{xy} = 2i$, and $\varepsilon_{yx} = -2i$.
(e)(f) $\varepsilon_{xx} = \varepsilon_{yy} = 9 - 3i$ and $\varepsilon_{xy} = \varepsilon_{yx} = 3 - i$.
(g)(h) $\varepsilon_{xx} = \varepsilon_{yy} = 9$ and $\varepsilon_{xy} = \varepsilon_{yx} = 2i$.
}
\label{fig:eigen}
\end{figure*}
In order to visualize some of conclusions we obtained by analytical equations in Sec. IIA, we present some numerical results of non-Hermitian anisotropic media.
We first consider a uniform reciprocal medium with $\varepsilon_{xx} = \varepsilon_{yy} = 9 - 3i$ and $\varepsilon_{xy}=\varepsilon_{yx} = 2$.
The corresponding $q$ is $0.2 + (1/15)i$.
Based on our previous results, this uniform medium should show NHSE.
The calculated eigenfrequency $\omega_{\text{PBC}}$ is plotted in Fig.~\ref{fig:eigen}(a).
The isofrequency contours of $\text{Re}(\omega_{\text{PBC}})$ and $\text{Im}(\omega_{\text{PBC}})$ form ellipsoids tilted from the $x$ and $y$ axes.
The inclination of the isofrequency contour reflects the shift of the wavevector caused by the real part of the effective gauge potential, and reflects the asymmmetry of the system in the $x$ direction.
Figure \ref{fig:eigen}(b) shows calculated $\omega_{\text{PBC}}$ and $\omega_{\text{PEC}}$ when $k_{y}L_{0}/2\pi = 0$ and $k_{y}L_{0}/2\pi = \pm 0.4$.
As predicted in Sec.~\ref{sec:real-k_theory}, the trajectory of $\omega_{\text{PBC}}$ and $\omega_{\text{PEC}}$ on the complex-$\omega$ plane are different when $k_{y} \neq 0$.
The trajectory of $\omega_{\text{PEC}}$ is inside the open arc drawn by $\omega_{\text{PBC}}$ when $k_{y} \neq 0$.
The shape of the trajectory of $\omega_{\text{PBC}}$ is the same at $k_{y}$ and $-k_{y}$.
However, the winding of the trajectory of $\omega_{\text{PBC}}$ when $k_{x}$ increases is opposite to each other: in Figs.~\ref{fig:eigen}(b) and (h), the trajectory winds clockwise when $k_{y}L_{0}/2\pi = 0.4$ while winds counterclockwise when $k_{y}L_{0}/2\pi = -0.4$.
As shown in Appendix \ref{sec:winding}, the direction of the winding is related to the winding number, and determined by the sign of $\text{Im}(qk_{y})$.
The amplitude of the electric field at $k_{y}L_{0}/2\pi = 0$ and $\pm0.4$ is also plotted in Fig.~\ref{fig:eigen}(b).
When $k_{y} = 0$, $|E_{y}|$ is extended over the system because of $q=0$.
On the other hand, $|E_{y}|$ is localized at the left boundary when $k_{y}L_{0}/2\pi = 0.4$ because of $\text{Im}(qk_{y}) > 0$. 
Similarly, $|E_{y}|$ is localized at the right boundary when $k_{y}L_{0}/2\pi = -0.4$ because of $\text{Im}(qk_{y}) < 0$. 
The localization length of the envelope of $|E_{y}|$ in Fig.~\ref{fig:eigen}(b) is accurately described by the analytical $|\text{Im}(qk_{y})|^{-1}$. Note that the position of the skin mode is determined by the winding direction, which is the same as the topological property of the NHSE in periodic systems.

As a second example, Figs.~\ref{fig:eigen}(c) and \ref{fig:eigen}(d) show the numerical results when the dielectric tensor is anti-symmetric: $\varepsilon_{xx} = 10 - 3i$, $\varepsilon_{xy} = 2i$, $\varepsilon_{yx} = -2i$, and $\varepsilon_{yy} = 8 - 3i$. 
Since the dielectric tensor is anti-symmetric, the corresponding $q$ is zero. Based on our previous analysis, this uniform media should not show NHSE.
In this case, the isofrequency contour of $\omega_{\text{PBC}}$ is symmetric with respect to $k_{x} = 0$ and $k_{y} = 0$ (Fig.~\ref{fig:eigen}(c)).
We can see that the trajectories of $\omega_{\text{PBC}}$ and $\omega_{\text{PEC}}$ agree even when $k_{y} \neq 0$ (Fig.~\ref{fig:eigen}(d)).
These numerical results visualize the disappearance of the NHSE in systems with anti-symmetric tensors.

As a third example, the NHSE in uniform media also vanishes when $\text{Im}(qk_{y}) = 0$.
Figure \ref{fig:eigen}(e) shows the isofrequency contour of $\omega_{\text{PBC}}$ when $\varepsilon_{xx} = \varepsilon_{yy} = 9 - 3i$ and $\varepsilon_{xy} = \varepsilon_{yx} = 3 - i$.
The two eigenvalues of the dielectric tensor are given by $\varepsilon_{1,2}$ = $12 - 4i$ and $6 - 2i$, and the corresponding $q$ is $1/3$.
Because of $\text{arg}(\varepsilon_{xx}) = \text{arg}(\varepsilon_{xy} + \varepsilon_{yx})$ (or $\text{Im}(\varepsilon_{1}\varepsilon_{2}^{*}) = 0$), $q$ becomes real.
The isofrequency contours of $\omega_{\text{PBC}}$ is tilted by the real effective gauge potential.
However, both $\text{Re}(\omega_{\text{PBC}})$ and $\text{Im}(\omega_{\text{PBC}})$ become symmetric with respect to $k_{x} + qk_{y} = 0$ when $q$ is real (see Eq.~(\ref{eq:PBC-omega})).
The trajectories of $\omega_{\text{PBC}}$ and $\omega_{\text{PEC}}$ agree even when $k_{y} \neq 0$ as shown in Fig.~\ref{fig:eigen}(f).
Thus, this medium does not show NHSE. In this particular case, 
although the system lacks the mirror symmetry with respect to the $yz$ plane, the NHSE accidentally vanishes when both $q$ and $k_{y}$ is real.  We note that $\text{Im}(q) = 0$, such as Fig.~\ref{fig:eigen}(e) and \ref{fig:eigen}(f), is a transition point between $\text{Im}(q)> 0 $ and $\text{Im}(q) < 0$.
By introducing perturbation of the anisotropy which switches the sign of $\text{Im}(q)$, we can switch the localization side of the skin mode if $k_{y}$ is real.

Finally, we discuss a system with the mirror-time symmetry, which corresponds to gain-loss balanced anisotropic media. Figures \ref{fig:eigen}(g) and \ref{fig:eigen}(h) plot the numerical results when $\varepsilon_{xx} = \varepsilon_{yy} = 9$ and $\varepsilon_{xy} = \varepsilon_{yx} = 2i$.
The corresponding $q$ is $2i/9$, and it becomes pure imaginary.
Based on our previous results, this medium should show NHSE.
Figure \ref{fig:eigen}(g) shows that $\text{Re}(\omega_{\text{PBC}})$ is symmetric with respect to $k_{x}=0$ and $k_{y} = 0$ and that $\text{Im}(\omega_{\text{PBC}})$ is anti-symmetric with respect to $k_{x}=0$ and $k_{y} = 0$.
The trajectory of $\omega_{\text{PEC}}$ is located inside the open arc drawn by $\omega_{\text{PBC}}$ when $k_{y} \neq 0$.
It should be noted that the skin mode has a real (or pure imaginary) $\omega_{\text{PEC}}$ in systems with the mirror-time symmetry because the mirror-time symmetry ensures the reality of $\omega_{\text{PEC}}^{2}$.

These numerical results visualize what type of anisotropy is required to show NHSE in media with gain or loss. The anisotropy should not have the mirror symmetry in the $x$- and $y$-axes, and the anisotropy of $\text{Re}(\omega)$ and $\text{Im}(\omega)$ should be different. Such anisotropy can be realized in uniaxially or biaxially anisotropic materials with gain or loss.

 \section{Non-Hermitian skin effect in multilayer metamaterials}
 \label{sec:3}
\subsection{Effective medium theory}
\begin{figure}
\includegraphics[width=8.6cm]{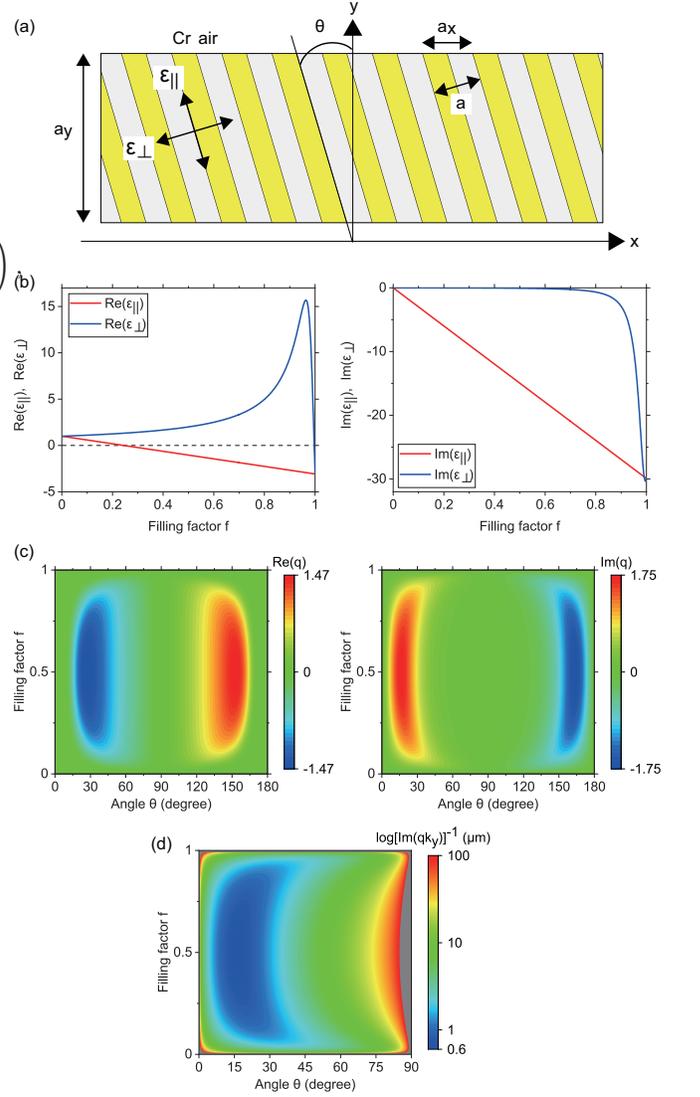}
\caption{
(a) Schematics of a tilted multilayer metamaterial with $f=0.5$ and $\theta=16.5^{\circ}$.
The period of the multilayer is denoted as $a$, and the tilted angle is denoted as $\theta$.
The multilayer is periodic in the $x$ direction and $y$ direction.
The period in the $x$ direction is given by $a_{x} = a/\cos\theta$, and the period in the $y$ direction is given by $a_{y} = a/\sin\theta$.
The structure is uniform in the $z$ direction.
(b) Numerical results of (left) $\varepsilon_{\parallel}$ and (right) $\varepsilon_{\perp}$ as a function of $f$.
(c) Numerical result of (left) $\text{Re}(q)$ and (right) $\text{Im}(q)$ as functions of $\theta$ and $f$.
(d) Numerical result of $|\text{Im}(qk_{y})|^{-1}$ at $k_{y} = 0.2(2\pi/\lambda_{0})$ as function of $\theta$ and $f$ with $\lambda_{0} = 1500$ nm.
The gray regions represents regions where $|\text{Im}(qk_{y})|^{-1}$ is over $100$ $\mu$m. 
}
\label{fig:multilayer}
\end{figure}

As explained in the introduction, we apply our framework of the NHSE in uniform media to metamaterials, which has subwavelength artificial structures.
To implement the anisotropy of a dielectric tensor, we here adopt multilayer metamaterials consisting of two alternating layers with subwavelength thickness.
We choose one layer as a dielectric layer, and the other layer as a metallic layer.
A metallic layer naturally leads to material loss, and a metal-insulator multilayer exhibits strong in-plane anisotropy~\cite{poddubny2013hyperbolic,narimanov2015naturally}.
The multilayer metamaterial is characterized by the period $a$ and filling factor of a metallic layer $f$.
When the operation wavelength is sufficiently long compared to the period of the multilayer, the multilayer can be regarded as a uniform material with an effective permittivity~\cite{poddubny2013hyperbolic}.
Here we define a longitudinal component of the effective permittivity $\varepsilon_{\parallel}$ and the transverse one $\varepsilon_{\perp}$ as shown in Fig.~\ref{fig:multilayer}(a).
They are described by the effective medium theory, and given by
\begin{align}
\varepsilon_{\parallel} = f\varepsilon_{1} + (1-f)\varepsilon_{2}, \quad \varepsilon_{\perp} = \frac{\varepsilon_{1}\varepsilon_{2}}{(1-f)\varepsilon_{1} + f\varepsilon_{2}}.
\label{eq:effective_medium_theory}
\end{align}
Note that this dielectric tensor has essentially the same form as that in uniaxially anisotropic crystals, and thus NHSE would be expected for this multilayer metamaterials.
In this paper, we consider a multilayer consisting of Cr and air.
The permittivity of Cr is assumed to be $\varepsilon_{\text{Cr}} = \varepsilon_{1} = -3.072 - 29.929i$, which is the value at $1500$ nm~\cite{PhysRevB.6.4370}.
The effective permittivity of the Cr-air multilayer is shown in Fig.~\ref{fig:multilayer}(b).
The effective permittivity satisfies $\text{Re}(\varepsilon_{\parallel}) < 0$ and $\text{Re}(\varepsilon_{\perp}) > 0$ in a wide range of $f$. 
The imaginary part of $\varepsilon_{\parallel}$ is larger than the imaginary part of $\varepsilon_{\perp}$ in a wide range of $f$.
The off-diagonal component of the dielectric tensor can be introduced by tilting the multilayer.
By rotating the coordinate system, we derive the effective dielectric tensor of the tilted multilayer metamaterial and the corresponding $q$ leading to the gauge potential contribution :
\begin{align}
\begin{pmatrix}
\varepsilon_{xx} && \varepsilon_{xy}
\\
\varepsilon_{yx} && \varepsilon_{yy}
\end{pmatrix}
=
\begin{pmatrix}
\varepsilon_{\perp}\cos^{2}\theta + \varepsilon_{\parallel}\sin^{2}\theta && \left( \varepsilon_{\perp} - \varepsilon_{\parallel} \right)\cos\theta\sin\theta
\\
\left( \varepsilon_{\perp} - \varepsilon_{\parallel} \right)\cos\theta\sin\theta && \varepsilon_{\parallel}\cos^{2}\theta + \varepsilon_{\perp}\sin^{2}\theta
\end{pmatrix},
\label{eq:effective_tensor}
\end{align}
\begin{align}
q = \frac{\left( \varepsilon_{\perp} - \varepsilon_{\parallel} \right) \cos\theta \sin\theta}{\varepsilon_{\perp} \cos^{2}\theta + \varepsilon_{\parallel} \sin^{2}\theta}.
\label{eq:q_effective}
\end{align}
Equation (\ref{eq:q_effective}) is consistent with Eq.~(\ref{eq:definition_q_2}) with $\delta = 0$.
We plot $q$ calculated from Eqs.~(\ref{eq:effective_medium_theory}) and (\ref{eq:q_effective}) as functions of $f$ and $\theta$ in Fig.~\ref{fig:multilayer}(c).
In the Cr-Air multilayer, the real and imaginary parts of the gauge potential parameter $q$ are generally finite.
Thus, this metamaterial should show NHSE based on our previous analysis.
The parameter $q$ can be tuned by the filling factor and angle of the multilayer.
The sign of $\text{Re}(q)$ and $\text{Im}(q)$ are different each other in this case.
The parameter $q$ is symmetric with respect to $f=0.5$, and antisymmetric with respect to $\theta = 90^{\circ}$.
The largest $\text{Im}(q)$ is achieved near $f = 0.5$ and $\theta = 16.5^{\circ}$.
At $f = 0.5$ and $\theta = 16.5^{\circ}$, the value of $q$ is $-0.7842 + 1.7407i$.
Figure \ref{fig:multilayer}(d) plots the corresponding localization length when $k_{y} = 0.2(2\pi/\lambda_{0})$ with $\lambda_{0} = 1500$ nm.
The localization length can be widely tuned by $f$ and $\theta$.
The localization length at $f=0.5$ and $\theta = 16.5^{\circ}$ is approximately estimated at $|\text{Im}(qk_{y})|^{-1} = 0.6857$ $\mu$m when $k_{y} = 0.2(2\pi/\lambda_{0})$.
 
\subsection{Numerical result with effective medium theory and finite element method}
\label{sec:multilayer_numerical}
\begin{figure*}
\includegraphics[width=17.2cm]{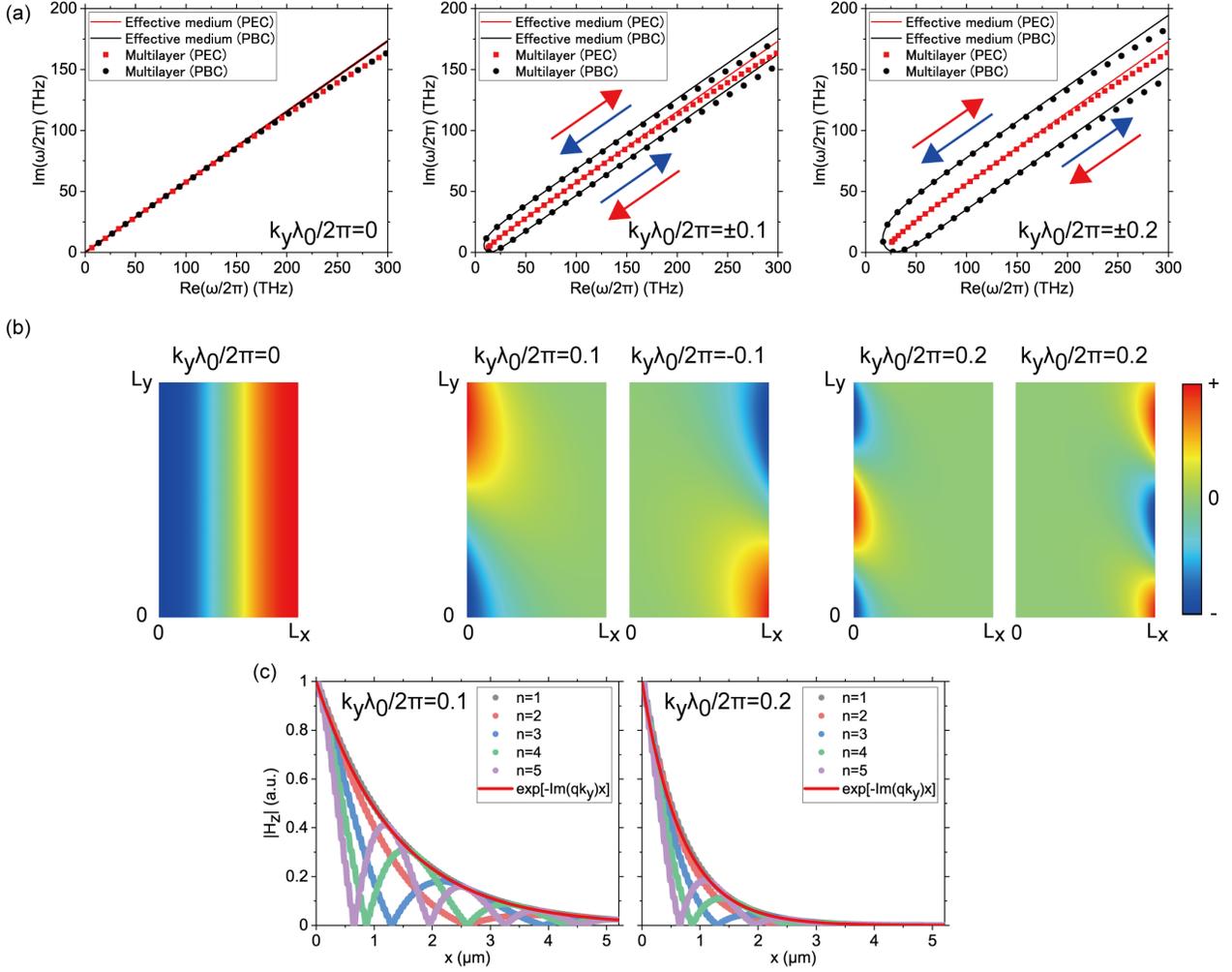}
\caption{
(a) Trajectory of the eigenfrequencies of the effective uniform medium and multilayer when $k_{y}\lambda_{0}/2\pi = 0$ (left), $\pm 0.1$ (middle), and $\pm 0.2$ (right), where $\lambda_{0} = 1500$ nm is the reference wavelength.
The period of the multilayer $a$ is set to 50 nm.
The size of the system in the $x$ direction is $L_{x} = 100 a_{x}$, where $a_{x} = a/\cos(\theta)$ represents the period of the multilayer in the $x$ direction.
Red and blue arrows represent the direction of the trajectory when $k_{x}$ increases.
Red arrows correspond to $k_{y} > 0$ and blue arrows correspond to $k_{y} < 0$.
(b) Numerical results of the eigenmode in the multilayer.
The magnetic field $H_{z}(x, y)$ of the lowest mode ($n=1$) under the PEC boundary condition at $k_{y}\lambda_{0}/2\pi = 0, \pm 0.1$, and $\pm 0.2$ are plotted.
We impose the Bloch boundary condition in the $y$ direction and $H_{z}(x,y)$ satisfies $H_{z}(x,y + a_{y}) = e^{-ik_{y}a_{y}}H_{z}(x, y)$, where $a_{y}= a/\sin(\theta)$ is the period of the multilayer in the $y$ direction.
The mode profile in the range from $y=0$ to $y=L_{y}$ is plotted with $L_{y}=500a_{y}$.
(c) Calculated $|H_{z}|$ of the multilayer at (left) $k_{y}\lambda_{0}/2\pi=0.1$ and (right) 0.2.
For comparison, $\text{exp}[-\text{Im}(qk_{y})]$ for (left) $k_{y}\lambda_{0}/2\pi = 0.1$ and (right) 0.2 is also plotted.
}
\label{fig:eigenmode}
\end{figure*}

In this section, we compare the result of a multilayer with that of an corresponding effective medium. 
We set $f$ and $\theta$ to $f=0.5$ and $\theta = 16.5^{\circ}$ because $\text{Im}(q)$ takes the maximum value around $\theta = 16.5^{\circ}$ and $f=0.5$.
The values of the corresponding effective dielectric tensor are $\varepsilon_{xx} = 1.759-1.268i$, $\varepsilon_{yy} = -0.791-13.763i$, and $\varepsilon_{xy} = \varepsilon_{yx} = 0.828+4.057i$.
The eigenfrequencies of the effective medium and the multilayer under the PBC and PEC boundary condition are plotted in Fig.~\ref{fig:eigenmode}(a).
The eigenfrequency of the multilayer is computed by using COMSOL Multiphysics.
In the calculation of the multilayer, the periodic boundary condition is imposed at $y=0$ and $y=a_{y}$, and the magnetic field in the multilayer satisfies the Bloch boundary condition $H_{z}(x,y+a_{y}) = e^{-ik_{y}y}H_{z}(x,y)$.
The eigenfrequencies $\omega_{\text{PBC}}$ and $\omega_{\text{PEC}}$ of the multilayer are in good agreement with those of the effective medium when the mode index $n$ is sufficiently small.
We can observe that the trajectory of $\omega_{\text{PEC}}$ is inside the trajectory of $\omega_{\text{PBC}}$ even in the multilayer.  
The numerical results for metamaterials start to deviate from those for analytical uniform media at high frequencies.
This deviation is reasonable because the effective medium theory should be invalid at high frequencies when the layer period is comparable or larger than the wavelength of light. At higher frequencies, metamaterials should be considered as one- or two-dimensional photonic crystals.
With this argument, the NHSE for uniform media should be adiabatically connected to the long-wavelength limit of the NHSE in the first band of photonic crystals~\cite{PhysRevB.104.125416,yan2021non,PhysRevResearch.4.023089,FangHuZhouDing+2022+3447+3456,PhysRevB.106.195412}.

The lowest eigenmodes with $n=1-5$ of the multilayer under the PEC boundary condition are plotted in Fig.~\ref{fig:eigenmode}(b).
The magnetic field $H_{z}(x,y)$ at $k_{y}\lambda_{0}/2\pi = 0$ is extended over the system, while $H_{z}(x,y)$ is localized at the left (right) boundary when $k_{y} > 0$ $(k_{y} < 0)$.
The numerical result of $|H_{z}(x)|$ of the multilayer is shown in Fig.~\ref{fig:eigenmode}(c).
The envelope of $|H_{z}(x)|$ in the multilayer decays exponentially in the $x$ direction, and the envelope agrees with $\text{exp}[-\text{Im}(qk_{y})]$ predicted by the effective medium theory.
The localization length in the $x$ direction $|\text{Im}(qk_{y})|^{-1}$ is estimated at $|\text{Im}(qk_{y})|^{-1} = 1.3741 \ \mu\text{m}$ when $k_{y}\lambda_{0}/2\pi = 0.1$ and $|\text{Im}(qk_{y})|^{-1} = 0.6857 \  \mu\text{m}$ when $k_{y}\lambda_{0}/2\pi = 0.2$.
The skin mode is strongly localized in a region with a size comparable to the wavelength.

The results presented in this section demonstrate that one can design metamaterials showing NHSE, which is essentially similar to the NHSE in uniform media. Importantly, various characteristics of NHSE can be tuned by controlling the parameters of multilayer metamaterials. The analytical framework enables us to design NHSE in versatile ways.

\section{Stationarily-excited mode in non-Hermitian anisotropic media}
\label{sec:4}
We have so far discussed the skin mode with a complex $\omega$ and a real $k_{y}$ in the non-Hermitian anisotropic media. These real $k_{y}$ skin modes are obtained as eigenfunctions of non-Hermitian systems. In fact, it is not trivial how to observe these eigenfunction skin modes because any optical observation requires an excitation process which may alter the mode profile in non-Hermitian systems. Thus, in order to clarify the observable NHSE, we investigate excited modes directly using the exact analytical formulation of the NHSE in anisotropic uniform media developed in the present work. 
In the previous section, $k_{y}$ is a parameter we can choose ($k_{y}$ is usually taken as real but $k_{y}$ can be extended to complex), and $\omega$ is determined by $k_{y}$ and $n$ via the dispersion relation.
Instead, in this section, in order to investigate stationary modes excited externally, we will take $\omega$ as a parameter ($\omega$ is taken as real but $\omega$ can be extented to complex~\cite{PhysRevLett.124.193901,gu2022transient}), and $k_{y}$ is determined by $\omega$ and $n$ via the dispersion relation.
Similar methods were taken in many text books of electromagnetism: for example, in Ref.~\cite{landau2013electrodynamics,jackson1999classical,pozar2011microwave}, the frequency of a mode in lossy media is assumed to be real.
Modes with real $\omega$ are suitable to describe experiments with a real-valued excitation frequency such as transmission and reflection measurements under stationary excitation.
Therefore, the analysis of modes with real $\omega$ is important for the experimental observability of the skin mode.
As explained in the introduction, it is important to take excitation processes into account to investigate NHSE observable in experiments.
We note that the response of the NHSE in one-dimensional systems under stationary excitation was investigated in Ref.~\cite{PhysRevResearch.2.013058,schomerus2022fundamental}, especially focusing the local density of states.
The present work deals with two-dimensional systems under stationary excitation, armed with the analytical formulation for anisotropic uniform media, and we are interested in mode profile, localization, and propagation. Our result reveals novel aspects of NHSE in this stationarily-excited situation, as we show below.

Before proceeding, we mention an experimental system we imagine, and how to excite the stationary skin mode.
To excite a real-$\omega$ skin mode, we should consider an interface between a lossless isotropic medium and a non-Hermitian anisotropic medium sandwiched by two PEC placed at $x=0$ and $x=L$.
Such discontinuity of the parallel-plate slab waveguide is usually analyzed by modal analysis~\cite{pozar2011microwave} based on an expansion with the transverse mode numbers. 
When a certain incident mode in the lossless isotropic region with an excitation frequency $\omega$ is injected to the non-Hermitian anisotropic medium, it is expected that an excited wave can be written by the sum of real-$\omega$ skin modes for different transverse mode numbers with the same frequency $\omega$.
Here, we consider a real-$\omega$ mode with a given $n$.
The general formulation of the modal analysis will be reported elsewhere.

\subsection{Theory}
\begin{figure*}
\includegraphics[width=17.2cm]{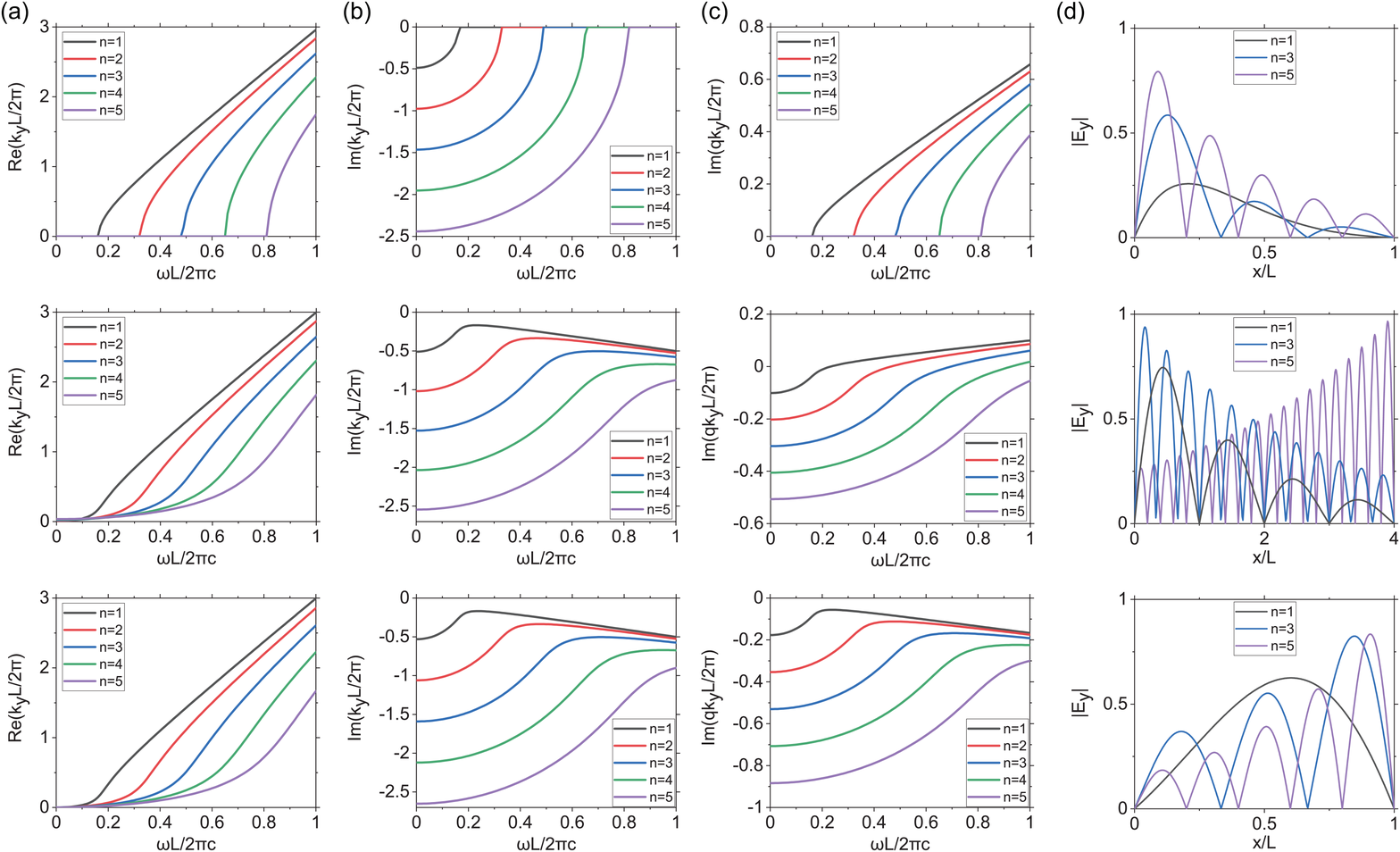}
\caption{
Numerical result of (a) $\text{Re}(k_{y})$, (b) $\text{Im}(k_{y})$, and (c) $\text{Im}(qk_{y})$ for $n=1-5$.
(d) Normalized $|E_{y}|$ for $n=1,3,5$.
The frequency is fixed at $\omega L/2\pi c =1$.
The dielectric tensor is set to $\varepsilon_{xx} = \varepsilon_{yy} = 9$ and $\varepsilon_{xy} = \varepsilon_{yx} = 2i$ in the upper panels, $\varepsilon_{xx} = \varepsilon_{yy} = 9-3i$ and $\varepsilon_{xy} = \varepsilon_{yx} = 2$ in the middle panels, and $\varepsilon_{xx} = \varepsilon_{yy} = 9-3i$ and $\varepsilon_{xy} = \varepsilon_{yx} = 3-i$ in the lower panels.
}
\label{fig:excited}
\end{figure*}
For a specific example, we reconsider a finite system of a non-Hermitian anisotropic medium.
In contrast to the previous section, we seek solutions with a real $\omega$.
The boundary condition at $x = 0$ and $L$ determines $k_{x}$, and $k_{y}$ is derived by solving the dispersion relation for $k_{y}$ as functions of $\omega$ and $k_{x}$.
Under the PEC boundary condition, $k_{\pm}$ and $k_{y}$ under the PEC boundary condition are given by
\begin{align}
k_{\pm} &= \pm \frac{\pi n}{L} - qk_{y, n}(\omega)
\\
k_{y, n}(\omega) 
&= \pm \sqrt{ \frac{1}{\eta_{xx} - q^{2}\eta_{yy}} \left\{ \left(\frac{\omega}{c}\right)^{2} - \eta_{yy}\left(\frac{\pi n}{L}\right)^{2} \right\} },
\label{eq:PEC-excite}
\end{align}
where $\omega$ is the real-valued excitation frequency.
The real-$\omega$ mode under the PEC boundary condition can be simply derived by inserting Eq.~(\ref{eq:PEC-excite}) into Eqs.~(\ref{eq:PEC-eigemode-Ey})-(\ref{eq:PEC-eigemode-Ex}).
Here $n$ is the transverse mode number for the finite-sized boundaries. Note that $k_{y, n}$ is determined through the dispersion relation for specific $n$, although $k_{y}$ is predetermined in the NHSE in the previous sections. 
This derivation shows that the stationarily-excited electromagnetic modes (that is, the real-$\omega$ modes) are expressed as the same equations as in the real-$k_{y}$ modes in the previous sections, and thus they are exponentially localized near one of the boundaries when $\text{Im}[qk_{y,n}(\omega)] \neq 0$.
Importantly, the localization of the real-$\omega$ mode is completely characterized by $\text{Im}(qk_{y})$, which is also the same as in the NHSE in the real-$k_{y}$ mode, and thus the localization of the real-$\omega$ mode is caused by the effective imaginary gauge potential as in the Hatano-Nelson model.
Therefore, we regard that these localized modes can be considered the NHSE in the stationary excited modes. 

The only difference between the real-$k_{y}$ mode and real-$\omega$ mode is which quantity ($k_{y}$ or $\omega$) is forced to be real.
This leads to some different characteristics between the stationarily-excited NHSE with real-$\omega$ modes and the conventional NHSE with real-$k_{y}$ modes.
First, $k_{y}$ depends on $\omega$ and $n$ for the real-$\omega$ mode.
Therefore, the localization length in the $x$ direction of real-$\omega$ skin modes depends on $\omega$ and $n$ via $k_{y,n}(\omega)$.
Second, $k_{y}$ is generally complex for the real-$\omega$ mode, and $\text{Im}[k_{y,n}(\omega)]$, which represents the attenuation constant in the $y$ direction, contributes to the localization of the real-$\omega$ skin mode because $\text{Im}(qk_{y})$ is written by $\text{Im}(qk_{y}) = \text{Im}(q)\text{Re}(k_{y}) + \text{Re}(q)\text{Im}(k_{y})$.
For the real-$k_{y}$ skin mode, only the first term contributes to $\text{Im}(qk_{y})$ because $k_{y}$ is real.
On the other hand, both the real and imaginary parts of $q$ play a role for the localization of the real-$\omega$ skin mode, as shown in Sec.~\ref{sec:stationary_numerical}.

In Sec.~\ref{sec:classification}, we derived conditions for the form of dielectric tensors that prohibit the real-$k_{y}$ modes.
Since the real-$\omega$ skin modes can be formed even for real $q$, the forbidden condition becomes more narrowed.
This means that the real-$\omega$ skin modes can be always formed for general anisotropic non-Hermitian media if we set the boundaries at an appropriate direction.

\subsection{Numerical result}
\label{sec:stationary_numerical}
\begin{figure}
\includegraphics[width=8.6cm]{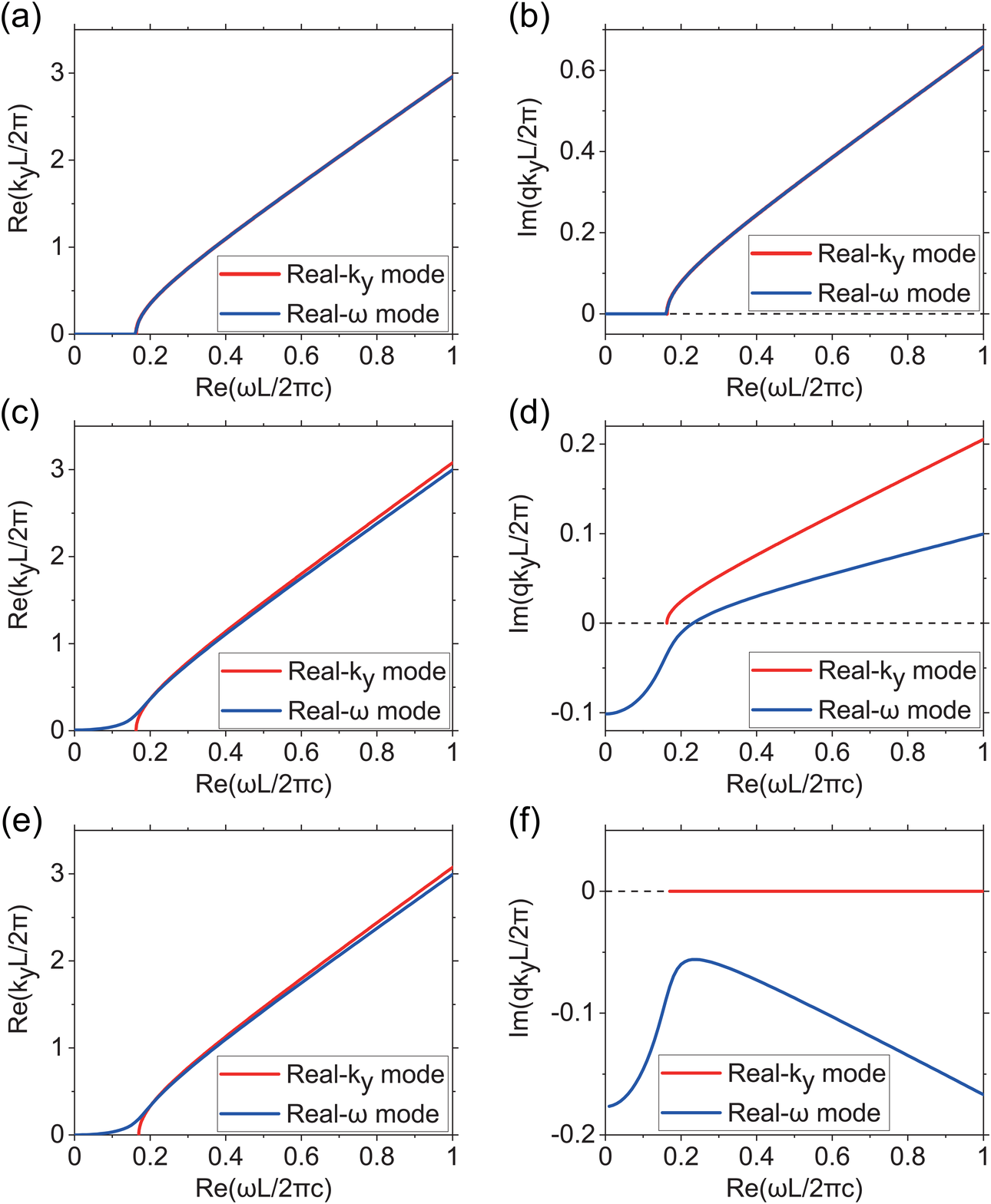}
\caption{
(a)(c)(e) Comparison of dispersion relations between $\text{Re}(k_{y})$ and $\text{Re}(\omega)$.
The dispersion relation is calculated from Eqs.~(\ref{eq:PEC-eigenfrequency}) and (\ref{eq:PEC-excite}).
(b)(d)(f) Comparison of the localization strength $\text{Im}(qk_{y})$ as a function of $\text{Re}(\omega)$.
The dielectric tensor is set to (a)(b) $\varepsilon_{xx} = \varepsilon_{yy} = 9$ and $\varepsilon_{xy} = \varepsilon_{yx} = 2i$, (c)(d) $\varepsilon_{xx} = \varepsilon_{yy} = 9 - 3i$ and $\varepsilon_{xy} = \varepsilon_{yx} = 2$, (e)(f) $\varepsilon_{xx} = \varepsilon_{yy} = 9 - 3i$ and $\varepsilon_{xy} = \varepsilon_{yx} = 3 - i$.
The mode index is fixed at $n=1$ in all plots.
}
\label{fig:excited_comparison}
\end{figure}
Here we present some numerical result of the real-$\omega$ skin mode, and compare it with the real-$k_{y}$ skin mode.
First, we begin with a special case with the mirror-time symmetry, where $\omega$ and $k_{y}$ can be simultaneously real although the system is non-Hermitian.
Figures \ref{fig:excited}(a) and \ref{fig:excited}(b) show the numerical result of $k_{y}$ as a function of $\omega$ for several mode index $n$.
Here the dielectric tensor is set to $\varepsilon_{xx} = \varepsilon_{yy} = 9$ and $\varepsilon_{xy} = \varepsilon_{yx} = 2i$.
In the mirror-time symmetric system, it is useful to define the cut-off frequency $\omega_{\text{c}}$ given by
\begin{align}
\frac{\omega_{\text{c}}}{c} = \sqrt{(\eta_{xx} - q^{2}\eta_{yy})k_{y}^{2}}.
\end{align}
As shown in Figs.~\ref{fig:excited}(a) and \ref{fig:excited}(b), $k_{y}$ is real (pure imaginary) above (below) the cut-off frequency if $\eta_{yy} > 0$.
Figure \ref{fig:excited}(c) plots the calculated $\text{Im}(qk_{y})$.
This shows that even for stationarily-exicted modes, $\text{Im}(qk_{y})$ becomes non-zero when real-valued $\omega$ is higher than the cut-off frequency.
The modes with $n=1,3,5$ are plotted in Fig.~\ref{fig:excited}(d).
We can confirm that the exponentially-decaying skin modes appear.
These two modes have different localization length because $k_{y}$ depends on $n$, in contrast to the real-$k_{y}$ skin mode.
Figures \ref{fig:excited_comparison}(a) and \ref{fig:excited_comparison}(b) show the comparison with the real-$\omega$ skin mode and real-$k_{y}$ skin mode.
In the case with the mirror-time symmetry, importantly, the dispersion and $\text{Im}(qk_{y})$ of real-$\omega$ skin modes completely coincides with those of real-$k_{y}$ skin modes.
We conclude that one can externally excite skin modes essentially the same as those in the previous sections, with the presence of the mirror-time symmetry.

Next, we investigate a more general situation in Fig.~\ref{fig:excited} and Figs.~\ref{fig:excited_comparison}(c)(d), where $\varepsilon_{xx} = \varepsilon_{yy} = 9-3i$ and $\varepsilon_{xy}=\varepsilon_{yx}=2$.
The imaginary part of $k_{y}$ is negative, and thus the real-$\omega$ mode is attenuated in the $y$ direction.
We observe that $\text{Im}(qk_{y})$ is non-zero in the wide frequency region, proving the existence of the NHSE in this case, as shown in Fig.~\ref{fig:excited} and Figs.~\ref{fig:excited_comparison}(c)(d).
However, the dispersion relation and $\text{Im}(qk_{y})$ of the real-$k_{y}$ mode and real-$\omega$ mode are now different each other.
We also observe that the sign of $\text{Im}(qk_{y})$ of the real-$\omega$ skin mode change while sweeping $\omega$ and $n$ (see Fig.~\ref{fig:excited}(d)).
This is because $\text{Re}(q)\text{Im}(k_{y})$ has the opposite sign of $\text{Im}(q)\text{Re}(k_{y})$.
The inversion of the sign involves the inversion of the localization position, and thus the present result demonstrates that the real-$\omega$ skin mode exhibit different localization in the $x$ direction compared to the real-$k_{y}$ mode.

Finally, we show another special case where $q$ is purely real ($\varepsilon_{xx} = \varepsilon_{yy} = 9-3i$ and $\varepsilon_{xy} = \varepsilon_{yx} = 3-i$) in Fig.~\ref{fig:excited} and Figs.~\ref{fig:excited_comparison}(e)(f).
Note that as shown previously in Fig.~\ref{fig:eigen}(e)(f), the real-$k_{y}$ mode does not exhibit NHSE when $q$ is purely real.
However, we observe that $\text{Im}(qk_{y})$ becomes non-zero in a wide frequency range for real-$\omega$ modes.
The present result shows that the localization strength $\text{Im}(qk_{y})$ of the real-$k_{y}$ mode vanishes for all $\text{Re}(\omega)$, while $\text{Im}(qk_{y})$ of the real-$\omega$ mode can become finite because of the imaginary part of $k_{y}$.

These results show that exponentially-decaying skin modes appear for satrionarily-excited cases (real-$\omega$ skin modes), which are caused by $\text{Im}(q)$ and $\text{Re}(q)$. The former generates very similar skin modes to those in eigenfunctions (real-$k_{y}$ skin modes), but the latter generates novel skin modes which do not have counterparts in real-$k_{y}$ skin modes. Interestingly, one can switch the localization position by manipulating these two different contributions, which may lead to novel aspects or applications of the NHSE.

\subsection{Estimation of propagation and localization length in multilayer metamaterial}
\begin{figure}
\includegraphics[width=8.6cm]{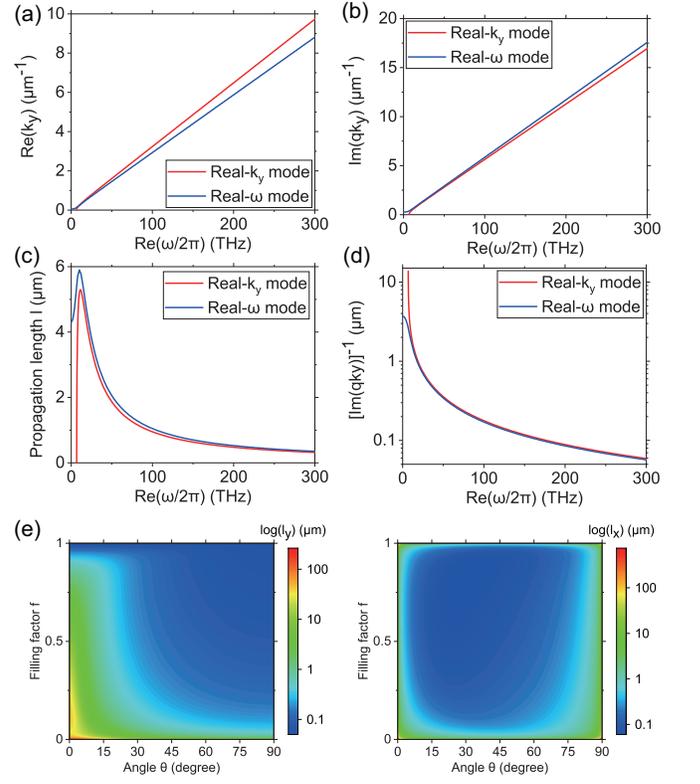}
\caption{
(a) Comparison of dispersion relation of the effective medium between $\text{Re}(k_{y})$ and $\text{Re}(\omega)$.
(b) Comparison of the localization strength $\text{Im}(qk_{y})$ as a function of $\text{Re}(\omega)$.
(c) Comparison of the propagation length in the $y$ direction calculated from real-$k_{y}$ mode and real-$\omega$ mode.
(d) Comparison of the localization length in the $y$ direction calculated from real-$k_{y}$ mode and real-$\omega$ mode.
(e) Numerical result of propagation length in the $y$ direction $l_{y}$ and (right) localization length in the $x$ direction $l_{x}$ as function of $f$ and $\theta$.
The excitation wavelength is fixed at $1500$ nm.
In all plots, the mode index is fixed at $n=1$.
}
\label{fig:excited_multilayer}
\end{figure}

We estimate the propagation length in the $y$ direction and localization length in the $x$ direction in the Cr-Air multilayer by using the effective medium theory.
To compare the real-$\omega$ skin mode with the real-$k_{y}$ skin mode discussed in Sec.~\ref{sec:multilayer_numerical}, we first set $f$ and $\theta$ to $f = 0.5$ and $\theta = 16.5^{\circ}$.
Figures \ref{fig:excited_multilayer}(a) and \ref{fig:excited_multilayer}(b) show the comparison of $\text{Re}(k_{y})$ and $\text{Im}(qk_{y})$ for the real-$k_{y}$ mode and real-$\omega$ mode with $n=1$.
Figure \ref{fig:excited_multilayer}(c) shows the propagation length in the $y$ direction.
For the real-$k_{y}$ mode, we define the propagation length as $\text{Im}(\omega_{\text{PEC}})^{-1}v_{g}$, where $v_{g} = d\text{Re}(\omega_{\text{PEC}})/dk_{y}$ is the group velocity of the skin mode.
For the real-$\omega$ mode, the propagation length is calculated by $|\text{Im}(k_{y})|^{-1}$.
When material loss or gain is sufficiently small, it is expected that the propagation length of the real-$k_{y}$ mode and real-$\omega$ mode coincides each other~\cite{doi:10.1021/acsphotonics.9b01202}.
In this case, however, the propagation length calculated from the two modes does not agree each other due to non-negligible material loss.
At $\text{Re}(\omega/2\pi) \approx 200$ THz, the propagation length of the real-$k_{y}$ mode is about $476$ nm, while the propagation length of the real-$\omega$ mode is about $527$ nm.
The numerical result of the localization length in the $x$ direction is plotted in Fig.~\ref{fig:excited_multilayer}(d).
In the effective medium of the multilayer, the only small difference appears.
At $\text{Re}(\omega/2\pi) \approx 200$ THz, the localization length of the real-$k_{y}$ mode is about $88.5$ nm, while the localization length of the real-$\omega$ mode is about $85.5$ nm.

Finally, we estimate the propagation length $l_{y}$ of the real-$\omega$ skin mode and localization length $l_{x}$ of the real-$\omega$ skin mode as functions of $f$ and $\theta$.
The excitation wavelength is fixed at $1500$ nm, and the mode index $n$ is fixed at $1$. 
The numerical result of the propagation length is shown in Fig.~\ref{fig:excited_multilayer}(e).
Smaller $f$ enhances the propagation length because the material loss can be reduced by reducing the volume of the metallic Cr layer.
In addition, smaller $\theta$ enhances the propagation length because $\varepsilon_{\perp}$ becomes more dominant than $\varepsilon_{\parallel}$ as $\theta$ becomes smaller.
In region where $f$ and $\theta$ are small, the propagation length reaches several $\mu$m.
Figure \ref{fig:excited_multilayer}(e) also shows the numerical calculation of the localization length.
For the real-$k_{y}$ mode, the strongest localization is achieved near $f=0.5$ and $\theta=16.5^{\circ}$ (Fig.~\ref{fig:multilayer}(c)), while the strongest localization is achieved near $f=0.82$ and $\theta = 29.5^{\circ}$ for the real-$\omega$ mode with excitation wavelength $1500$ nm and $n=1$.
This deviation is caused by the imaginary part of $k_{y}$.
The localization length of the real-$\omega$ mode is less than $1\ \mu$m in the wide range of $f$ and $\theta$, which indicates that the real-$\omega$ skin mode is localized in a region as small as the operation wavelength.

\section{Conclusion}
In conclusion, we have theoretically demonstrated that TE modes in non-Hermitian anisotropic media exhibits the NHSE.
The NHSE occurs when $\text{Im}(qk_{y}) \neq 0$, and its localization length in the $x$ direction is given by $|\text{Im}(qk_{y})|^{-1}$.
A skin mode propagating in the $+y$ direction is localized at a boundary of a system, while the counter-propagating skin mode is localized at the opposite boundary.
This peculiar localization arises from the combination of the non-Hermiticity and anisotropy.
At a glance, it seems that this phenomenon is non-reciprocal like topological edge modes in photonic Chern insulators without time-reversal symmetry~\cite{PhysRevLett.100.013904,PhysRevA.78.033834,PhysRevLett.100.013905,wang2009observation}.
However, the propagation of the skin mode discussed in this paper is essentially reciprocal because the NHSE occurs even when $\varepsilon$ is symmetric.
The advantage of using uniform media is that the NHSE in uniform media can be analytically predicted.
The NHSE also occurs in photonic crystals with appropriate structure and dielectric permittivity.
However, we cannot predict the strength of the localization in photonic crystals without detailed numerical calculations.
On the other hand, the NHSE in uniform media is completely governed by a dielectric tensor.
Because a dielectric tensor can be tuned by external fields, the NHSE in uniform media also may be controlled by external fields.
We also have proposed a new concept of a stationarily-excited skin mode.
Interestingly, in non-Hermitian anisotropic media, the spatial distribution of an eigenmode differs from that of an excited mode in contrast to Hermitian systems.
The notion of a stationarily-excited skin mode can be extended to two-dimensional periodic crystals although methods to calculate solutions with real $\omega$ and complex $k_{y}$ in periodic systems have not been established to our knowledge.
Our theory brings the simplest model of NHSE in two-dimensional systems, and it is useful for a better understanding of NHSE.
The theory developed in this paper can be extended to other classical wave systems.
Our work also pave the way to realize the optical NHSE in bulk materials such as metamaterials.

\begin{acknowledgments}
This work was supported by JSPS KAKENHI Grant Numbers JP20H05641, JST PRESTO Grant Number JPMJPR18L9 Japan, JSPS KAKENHI Grant Number JP21K14551, MEXT initiative to Establish Next-generation Novel integrated Circuits Centers (X-NICS) Grant Number JPJ011438, JSPS KAKENHI Grant Number JP22K18687 and JP22H00108.
K. Y. acknowledgements support from JSPS KAKENHI through Grant No. JP21J01409.
\end{acknowledgments}

\appendix
\section{Relation of eigenfrequney between PBC and PEC boundary condition}
\label{sec:PBC-PEC}
\begin{figure}
\includegraphics[width=8.6cm]{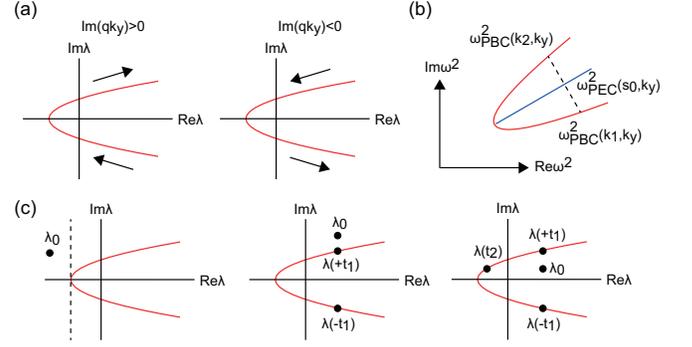}
\caption{
(a) Schematic illustration of parabolas.
Black arrows represents the direction of parabolas when $t$ increases.
(b) Schematic illustration of the trajectory of $\omega_{\text{PBC}}^{2}$ and $\omega_{\text{PEC}}^{2}$ on the complex-$\omega^{2}$ plane.
Red (blue) line represents the trajectory of $\omega_{\text{PBC}}^{2}$ ($\omega_{\text{PEC}}^{2}$). 
(c) Schematic illustration of various positions of $\lambda_{0}$.
Broken line represents $\text{Re}(\lambda) = -|\eta_{yy}| \{ \text{Im}(qk_{y}) \}^{2}$.
In the left and middle panels, $\lambda_{0}$ is located outside a parabola, whie $\lambda_{0}$ is located inside a parabola in the right panel.
}
\label{fig:parabola}
\end{figure}

\subsection{Trajectory of $\omega_{\text{PBC}}$ and $\omega_{\text{PEC}}$ on complex $\omega^{2}$ plane}
In the limit of $L \rightarrow \infty$, $\omega_{\text{PEC}}$ is expressed by
\begin{align}
\left( \frac{\omega_{\text{PEC}}}{c} \right)^{2}
= \eta_{yy}s^{2} + \left( \eta_{xx} - q^{2}\eta_{yy} \right)k_{y}^{2},
\end{align}
where $s$ is the positive real number.
For simplicity, we define $\lambda_{\text{PEC}} = (\omega_{\text{PEC}}/c)^{2} - (\eta_{xx} - q^{2}\eta_{yy})k_{y}^{2}$.
The real and imaginary parts of $\lambda_{\text{PEC}}$ are given by
\begin{align}
\text{Re} \left( \lambda_{\text{PEC}} \right) = \text{Re}\left( \eta_{yy} \right)s^{2},
\\
\text{Im} \left( \lambda_{\text{PEC}} \right) = \text{Im} \left( \eta_{yy} \right)s^{2}.
\end{align}
By eliminating $s$, we obtain
\begin{align}
\text{Im} \left( \lambda_{\text{PEC}} \right)
= \frac{\text{Im} \left( \eta_{yy} \right)}{\text{Re} \left( \eta_{yy} \right)} \text{Re} \left( \lambda_{\text{PEC}} \right).
\end{align}
Therefore, the trajectory of $\omega_{\text{PEC}}^{2}$ is a semi-infinite line on the complex-$\omega^{2}$ plane, and its inclination is given by $\text{Im}(\eta_{yy})/\text{Re}(\eta_{yy})$.

Similarly, in the limit of $L \rightarrow \infty$, $\omega_{\text{PBC}}$ is given by
\begin{align}
\left( \frac{\omega_{\text{PBC}}}{c} \right)^{2}
= \eta_{yy} \left( k_{x} + qk_{y} \right)^{2} + \left( \eta_{xx} - q^{2}\eta_{yy} \right)k_{y}^{2},
\end{align}
where $k_{x}$ is the real number.
Here we define $\lambda_{\text{PBC}} = (\omega_{\text{PBC}}/c)^{2} - (\eta_{xx} - q^{2}\eta_{yy})k_{y}^{2}$ for simplicity.
Defining $t = k_{x} + \text{Re}(qk_{y})$ yields
\begin{align}
\lambda_{\text{PBC}}
= \eta_{yy} \left[ t + i\text{Im} \left( qk_{y} \right) \right]^{2}.
\end{align}
The real and imaginary parts of $\lambda_{\text{PBC}}$ are given by
\begin{align}
\text{Re} \left( \lambda_{\text{PBC}} \right) &= \text{Re}(\eta_{yy}) \left[ t^{2} - \left\{ \text{Im}\left( qk_{y} \right) \right\}^{2} \right]
\nonumber \\
&- 2\text{Im} \left( \eta_{yy} \right) \text{Im}(qk_{y})t,
\\
\text{Im} \left( \lambda_{\text{PBC}} \right) &= \text{Im}(\eta_{yy}) \left[ t^{2} - \left\{ \text{Im}\left( qk_{y} \right) \right\}^{2} \right]
\nonumber \\
&+ 2\text{Re}\left( \eta_{yy} \right)\text{Im}\left( qk_{y} \right)t,
\end{align}
Next, we perform a rotational coordinate transformation.
We define $\lambda_{1}$ and $\lambda_{2}$ as
\begin{align}
\begin{pmatrix}
    \lambda_{1} \\ \lambda_{2}
\end{pmatrix}
= \frac{1}{\left| \eta_{yy} \right|}
\begin{pmatrix}
    \text{Re} \left( \eta_{yy} \right) && \text{Im} \left( \eta_{yy} \right)
    \\
    -\text{Im} \left( \eta_{yy} \right) && \text{Re} \left( \eta_{yy} \right)
\end{pmatrix}
\begin{pmatrix}
    \text{Re} \left( \lambda_{\text{PBC}} \right) \\ \text{Im} \left( \lambda_{\text{PBC}} \right)
\end{pmatrix}.
\end{align}
$\lambda_{1,2}$ are calculated as
\begin{align}
&\lambda_{1} = \left| \eta_{yy} \right| \left[ t^{2} - \left\{ \text{Im} \left( qk_{y} \right) \right\}^{2} \right],
\\
&\lambda_{2} = 2 \left| \eta_{yy} \right| \text{Im} \left( qk_{y} \right) t
\label{eq:lambda1_2}
\end{align}
Eliminating $t$, we can derive the relation between $\lambda_{1}$ and $\lambda_{2}$ for $\text{Im}(qk_{y}) \neq 0$,
\begin{align}
\lambda_{1} = \frac{\lambda_{2}^{2}}{4 \left| \eta_{yy} \right| \left\{ \text{Im} \left( qk_{y} \right) \right\}^{2}}
- \left| \eta_{yy} \right| \left\{ \text{Im} \left( qk_{y} \right) \right\}^{2}.
\end{align}
Therefore, the trajectory of $\omega_{\text{PBC}}^{2}$ is a parabola when $\text{Im}(qk_{y}) \neq 0$, while the trajectory of $\omega_{\text{PBC}}$ becomes a semi-infinite line when $\text{Im}(qk_{y}) = 0$.
The sign of $\text{Im}(qk_{y})$ determines the direction of the trajectory of a parabola when $t$ increases, as illustrated in Fig.~\ref{fig:parabola}(a).

\subsection{Middle-point theorem}
In this section, we will prove that the trajectory of $\omega_{\text{PEC}}^{2}$ is inside an open arc drawn by $\omega_{\text{PBC}}^{2}$.
Let us consider $\omega_{\text{PBC}}^{2}$ at $k_{x} = k_{1,2} = \pm \sqrt{s_{0}^{2} + \{\text{Im}(qk_{y})\}^{2}} - \text{Re}(qk_{y})$ with a positive real $s_{0}$.
They are given by
\begin{align}
&\left( \frac{\omega_{\text{PBC}} \left( k_{1,2}, k_{y} \right)}{c} \right)^{2}
\nonumber \\
&= \eta_{yy} \left[ s_{0}^{2} \pm 2i \text{Im} \left( qk_{y} \right) \sqrt{s_{0}^{2} + \left\{ \text{Im} \left( qk_{y} \right) \right\}^{2}} \right]
\nonumber \\
&+ \left( \eta_{xx} - q^{2}\eta_{yy} \right)k_{y}^{2}.
\end{align}
The average of them is
\begin{align}
&\frac{1}{2} \left[ \left( \frac{\omega_{\text{PBC}} \left( k_{1}, k_{y} \right)}{c} \right)^{2}
+ \left( \frac{\omega_{\text{PBC}} \left( k_{2}, k_{y} \right)}{c} \right)^{2} \right]
\nonumber \\
= &\eta_{yy}s_{0}^{2} + \left( \eta_{xx} - q^{2}\eta_{yy} \right)k_{y}^{2}
= \left( \frac{\omega_{\text{PEC}} \left( s_{0}, k_{y} \right)}{c} \right)^{2}.
\label{eq:intermediate}
\end{align}
Equation (\ref{eq:intermediate}) shows that $\omega_{\text{PEC}}(s_{0}, k_{y})$ lies at the midpoint between $\omega_{\text{PBC}}(k_{1}, k_{y})$ and $\omega_{\text{PBC}}(k_{2}, k_{y})$ on the complex $\omega^{2}$ plane as illustrated in Fig.~\ref{fig:parabola}(b).
If $\text{Im}(qk_{y}) = 0$, $\omega^{2}_{\text{PBC}}(k_{1}, k_{y})$, $\omega^{2}_{\text{PBC}}(k_{2}, k_{y})$, and $\omega^{2}_{\text{PEC}}(s_{0}, k_{y})$ satisfy
\begin{align}
\omega^{2}_{\text{PBC}} \left( k_{1}, k_{y} \right)
= \omega^{2}_{\text{PBC}} \left( k_{2}, k_{y} \right)
= \omega^{2}_{\text{PEC}} \left( s_{0}, k_{y} \right).
\end{align}

\section{Eigenmode analysis under PMC boundary condition}
\label{sec:PMC}
We consider a system sandwiched by two perfect magnetic conductors (PMCs) placed at $x=0$ and $x=L$.
The PMCs require $H_{z}(0) = H_{z}(L) = 0$, which gives
\begin{align}
\begin{pmatrix}
1 && 1
\\
e^{-ik_{-}L} && e^{-ik_{+}L}
\end{pmatrix}
\begin{pmatrix}
A \\ B
\end{pmatrix}
= 0.
\label{eq:PMC}
\end{align}
The nontrivial solution of Eq.~(\ref{eq:PMC}) is derived when $k_{+} - k_{-} = 2\pi n/L$.
Therefore, $k_{\pm}$ and eigenfrequency under the PMC boudary condition is same as those under the PEC boundary condition.
The eigenmode is given by
\begin{align}
H_{z}(x) &= e^{iqk_{y}x} \sin\left( \frac{\pi n}{L}x \right),
\label{eq:PMC-eigenmode-Hz}
\\
E_{x}(x) &= -\frac{1}{2}ie^{iqk_{y}x}
\Biggl[
\left( Z_{y+} - Z_{y-} \right) \cos\left( \frac{\pi n}{L}x \right)
\nonumber \\
&-i \left( Z_{y+} + Z_{y-} \right) \sin\left( \frac{\pi n}{L}L \right)
\Biggr],
\label{eq:PMC-eigenmode-Ex}
\\
E_{y}(x) &= \frac{1}{2}ie^{iqk_{y}x}
\Biggl[
\left( Z_{x+} - Z_{x-} \right) \cos\left( \frac{\pi n}{L}x \right)
\nonumber \\
&-i \left( Z_{x+} + Z_{x-} \right)\sin\left( \frac{\pi n}{L}x \right)
\Biggr].
\label{eq:PMC-eigenmode-Ey}
\end{align}
The envelope of the eigenmode (\ref{eq:PMC-eigenmode-Hz})-(\ref{eq:PMC-eigenmode-Ey}) exhibits exponential decay in the $x$ direction when $\text{Im}(qk_{y}) \neq 0$.

\section{Mirror-time symmetry}
\label{sec:mirror-time}
Mirror-time operation is the combination of a mirror operation and the time-reversal symmetry
We first consider a mirror operation with respect to the $xz$ plane denoted by $M_{y}$.
Because the $M_{y}$ operation transforms $(E_{x}, E_{y})$ into $(-E_{x}, E_{y})$, the $M_{x}$ operation transforms $\varepsilon$ as
\begin{align}
\varepsilon' = 
\begin{pmatrix}
\varepsilon_{xx} && -\varepsilon_{xy}
\\
-\varepsilon_{yx} && \varepsilon_{yy}
\end{pmatrix}.
\label{eq:mirror}
\end{align}
The time-reversal operation is represented by complex conjugation, and transforms $\varepsilon$ as $\varepsilon' = \varepsilon^{*}$.

The mirror-time ($M_{y}T$) operation transforms $\varepsilon$ as
\begin{align}
\varepsilon' =
\begin{pmatrix}
\varepsilon_{xx}^{*} && -\varepsilon_{xy}^{*}
\\
-\varepsilon_{yx}^{*} && \varepsilon_{yy}^{*}
\end{pmatrix}.
\end{align}
The $M_{y}T$ symmetry for $\varepsilon$ is represented by $\varepsilon' = \varepsilon$.
A system is invariant under the $M_{y}T$ symmetry when the diagonal components are real and off-diagonal components are pure imaginary.
The $M_{y}T$ operation transforms $\hat{\Theta}(k_{y})$ as $\hat{\Theta}'(k_{y}) = (M_{y}T)^{-1}\hat{\Theta}(k_{y})(M_{y}T)$, where $\hat{\Theta}'(k_{y})$ can be derive by complex conjugation as
\begin{align}
\hat{\Theta}'(k_{y}) = -\eta_{yy}^{*}\frac{d^{2}}{dx^{2}} + ik_{y}(\eta_{xy}^{*} + \eta_{yx}^{*}) \frac{d}{dx} + \eta_{xx}^{*}k_{y}^{2}.
\end{align}
When $\varepsilon' = \varepsilon$, $\hat{\Theta}(k_{y})$ satisfies the $M_{y}T$ symmetry represented by $\hat{\Theta}(k_{y}) = (M_{y}T)
^{-1} \hat{\Theta}(k_{y}) (M_{y}T)$.
The $M_{y}T$ symmetry leads to the reality of $\omega^{2}$ although $\hat{\Theta}(k_{y})$ is non-Hermitian.
In $M_{y}T$-symmeric systems, skin modes have real eigenfrequencies.

\section{Calculation of winding number}
\label{sec:winding}
Here we define the topological winding number of $\omega_{\text{PBC}}^{2}$.
To characterize the topological nature of a system, we use $\omega_{\text{PBC}}^{2}$ instead of $\omega_{\text{PBC}}$ because $\omega_{\text{PBC}}^{2}$ is the eigenvalue of $\hat{\Theta}(k_{y})$.
Because the winding number does not change under the coordinate rotation and the shift of the origin, we consider the complex function $\lambda(t) = \lambda_{1} + i\lambda_{2}$ defined by Eq.~(\ref{eq:lambda1_2}) without loss generality.
The winding number is defined by~\cite{PhysRevB.104.125109}
\begin{align}
W \left( \lambda_{0} \right) &= \frac{1}{2\pi} \int_{-\infty}^{\infty} dt \frac{d}{dt} \text{arg} \left[ \lambda(t) - \lambda_{0} \right],
\end{align}
where $\lambda_{0}$ is a reference point on the complex plane.

The winding number can be analytically calculated.
Here we define $\Lambda(t)=[\lambda_{2}(t) - \text{Im}(\lambda_{0})]/[\lambda_{1}(t) - \text{Re}(\lambda_{0})]$ for simplicity.
$\Lambda(t)$ converges to zero in the limit of $t\rightarrow \pm \infty$.
We first consider a case when $\text{Re}(\lambda_{0}) < -|\eta_{yy}|\{ \text{Im}(qk_{y}) \}^{2}$ (see the left panel of Fig.~\ref{fig:parabola}(c)).
In this case, $\lambda_{1}(t) - \text{Re}(\lambda_{0})$ satisfies $\lambda_{1}(t) - \text{Re}(\lambda_{0}) > 0$ for all $t$.
Therefore, the winding number is easily calculated as
\begin{align}
W \left( \lambda_{0} \right)
= \frac{1}{2\pi} \left[ \tan^{-1} \Lambda \left( t \right) \right]^{t=+\infty}_{t=-\infty} = 0.
\end{align}

Next, let us consider a case when $\lambda_{0}$ satisfies $\text{Re}(\lambda_{0}) > -|\eta_{yy}|\{\text{Im}(qk_{y})\}^{2}$ and $\text{Im}(\lambda_{0}) > 2|\eta_{yy}||\text{Im}(qk_{y})|t_{1}$ with
\begin{align}
t_{1} = \sqrt{ \{ \text{Im}(qk_{y}) \}^{2} + \frac{1}{\left| \eta_{yy} \right|} \text{Re}\left( \lambda_{0} \right) },
\end{align}
as illustrated in the middle panel of Fig.~\ref{fig:parabola}(c).
In this case, the winding number is calculated as
\begin{align}
W \left( \lambda_{0} \right)
&= \frac{1}{2\pi} \left[ \tan^{-1} \Lambda \left( t \right) \right]^{t=-t_{1}}_{t=-\infty}
+ \frac{1}{2\pi} \left[ \tan^{-1} \Lambda \left( t \right) - \pi \right]^{t=t_{1}}_{t=-t_{1}}
\nonumber \\
&+ \frac{1}{2\pi} \left[ \tan^{-1} \Lambda \left( t \right) \right]^{t=+\infty}_{t=t_{1}}
\nonumber \\
&= 0.
\end{align}
Similarly, the winding number when $\lambda_{0}$ satisfies $\text{Re}(\lambda_{0}) > -|\eta_{yy}|\{\text{Im}(qk_{y})\}^{2}$ and $\text{Im}(\lambda_{0}) < -2|\eta_{yy}||\text{Im}(qk_{y})|t_{1}$ is calculate as
\begin{align}
W \left( \lambda_{0} \right)
&= \frac{1}{2\pi} \left[ \tan^{-1} \Lambda \left( t \right) \right]^{t=-t_{1}}_{t=-\infty}
+ \frac{1}{2\pi} \left[ \tan^{-1} \Lambda \left( t \right) + \pi \right]^{t=t_{1}}_{t=-t_{1}}
\nonumber \\
&+ \frac{1}{2\pi} \left[ \tan^{-1} \Lambda \left( t \right) \right]^{t=+\infty}_{t=t_{1}}
\nonumber \\
&= 0.
\end{align}

Finally, we consider a case when  $\lambda_{0}$ satisfies $\text{Re}(\lambda_{0}) > -|\eta_{yy}|\{\text{Im}(qk_{y})\}^{2}$ and $|\text{Im}(\lambda_{0})| < 2|\eta_{yy}||\text{Im}(qk_{y})|t_{1}$
In this case, $\lambda_{0}$ is located inside a parabola, as illustrated in the right panel of Fig~\ref{fig:parabola}(c).
Here we define $t_{2}$ as
\begin{align}
t_{2} = \frac{\text{Im}\left( \lambda_{0} \right)}{2 \left| \eta_{yy} \right| \text{Im} \left( qk_{y} \right)},
\end{align}
where $t_{2}$ satisfies $\lambda_{2}(t_{2}) = \text{Im}(\lambda_{0})$ and $-t_{1} < t_{2} < t_{1}$.
The winding number is calculated as
\begin{align}
W \left( \lambda_{0} \right)
&= \frac{1}{2\pi} \left[ \tan^{-1} \Lambda \left( t \right) \right]^{t=-t_{1}}_{t=-\infty}
\nonumber \\
&+ \frac{1}{2\pi} \left[ \tan^{-1} \Lambda \left( t \right) - \text{sgn} \left\{ \text{Im}\left( qk_{y} \right) \right\} \pi \right]^{t=t_{2}}_{t=-t_{1}}
\nonumber \\
&+ \frac{1}{2\pi} \left[ \tan^{-1} \Lambda \left( t \right) + \text{sgn} \left\{ \text{Im}\left( qk_{y} \right) \right\} \pi \right]^{t=t_{1}}_{t=t_{2}}
\nonumber \\
&+ \frac{1}{2\pi} \left[ \tan^{-1} \Lambda \left( t \right) \right]^{t=+\infty}_{t=t_{1}}
\\
&= -\text{sgn} \left[ \text{Im}\left( qk_{y} \right) \right].
\end{align}
In summary, the winding number $W(\lambda_{0})$ is quantized to integers, and the winding number becomes $\pm 1$ if a reference point $\lambda_{0}$ is inside a parabola.

\section{Left eigenmode of Eq.~(\ref{eq:eigenequation})}
The transpose of a differential operator is defined by integration by parts~\cite{siegman1986lasers}.
The transpose of $\hat{\Theta}(k_{y})$ is given by~\cite{PhysRevB.104.125416}
\begin{align}
\hat{\Theta}^{T}(k_{y})
&= -\eta_{yy}\frac{d^{2}}{dx^{2}} + 2ik_{y}(\eta_{xy} + \eta_{yx})\frac{d}{dx} + k_{y}^{2}\eta_{xx}
\nonumber \\
&= \hat{\Theta}(-k_{y}).
\label{eq:Theta_transpose}
\end{align}
Equation (\ref{eq:Theta_transpose}) shows that the left eigenmode of $\hat{\Theta}(k_{y})$ is just the right eigenmode of $\hat{\Theta}(-k_{y})$.
The left eigenmode of $\hat{\Theta}(k_{y})$ under the PEC boundary condition is given by
\begin{align}
E^{L}_{y,k_{y},n}(x) &= e^{-iqk_{y}x}\sin\left( \frac{\pi n}{L}x \right),
\\
H^{L}_{z,k_{y},n}(x) &= \frac{1}{2}ie^{-iqk_{y}x}
\Biggl[ \left( \frac{1}{Z_{x+}} - \frac{1}{Z_{x-}} \right) \cos\left( \frac{\pi n}{L}x \right) 
\nonumber \\
&- i \left( \frac{1}{Z_{x+}} + \frac{1}{Z_{x-}} \right) \sin\left( \frac{\pi n}{L}x \right) \Biggr],
\\
E^{L}_{x,k_{y},n}(x) &= -\frac{1}{2}ie^{-iqk_{y}x}
\Biggl[ \left( \frac{Z_{y+}}{Z_{x+}} - \frac{Z_{y-}}{Z_{x-}} \right) \cos\left( \frac{\pi n}{L}x \right)
\nonumber \\
&- i \left( \frac{Z_{y+}}{Z_{x+}} + \frac{Z_{y+}}{Z_{x-}} \right) \sin\left( \frac{\pi n}{L}x \right) \Biggr],
\end{align}
The left eigenmode is localized at the opposite side to the right eigenmode.
We can easily prove the orthogonal relation by using the orthogonality of sine and cosine:
\begin{align}
\int_{0}^{L}H^{L}_{z,mk_{y}}(x) H^{R}_{z,nk_{y}}(x)dx \propto \delta_{mn}.
\end{align}

\bibliography{refs}

\end{document}